\documentclass[prd,aps,twocolumn,a4paper,floatfix,showpacs,nofootinbib]{revtex4-1}

\usepackage[utf8]{inputenc}  
\usepackage{lineno}
\usepackage[T1]{fontenc}
\usepackage{graphicx,psfrag}
\graphicspath{{figures/}}
\usepackage{mathrsfs}
\usepackage{amsmath,amsfonts,amssymb}
\usepackage{multirow,enumerate}
\usepackage{comment,hyperref}
\usepackage{color}
\usepackage{acronym}
\usepackage{algpseudocode}
\usepackage{bm}
\usepackage{lipsum}
\usepackage{algorithm}

\usepackage{mathtools}
\usepackage{cuted}
\usepackage{alltt}

\usepackage{array}
\newcolumntype{L}[1]{>{\raggedright\let\newline\\\arraybackslash\hspace{0pt}}m{#1}}
\newcolumntype{C}[1]{>{\centering\let\newline\\\arraybackslash\hspace{0pt}}m{#1}}
\newcolumntype{R}[1]{>{\raggedleft\let\newline\\\arraybackslash\hspace{0pt}}m{#1}}

\usepackage{xspace}
\usepackage[normalem]{ulem}
\usepackage{mathtools}
\usepackage{subfigure}
\usepackage{}
\usepackage{makecell}
\usepackage{color}

\usepackage{appendix}
\usepackage{diagbox}
\usepackage{lineno}

\newacro{BH}{black hole}
\newacro{NS}{neutron star}
\newacro{PN}{Post-Newtonian}
\newacro{BBH}{binary black hole}
\newacro{BNS}{binary neutron star}
\newacro{EOB}{effective-one-body}
\newacro{NR}{numerical relativity}
\newacro{GW}{gravitational wave}
\newacro{EOS}{equation-of-state}

\newcommand{\be}{\begin{equation}}
\newcommand{\ee}{\end{equation}}
\newcommand{\bea}{\begin{eqnarray}}
\newcommand{\eea}{\end{eqnarray}}
\newcommand{\bel}{\begin{align}}
\newcommand{\eel}{\end{align}}

\def\GMc2{{\rm G M_{\odot} c^{-2}}}

\def\SEOBNRv4T{\texttt{SEOBNRv4T}\xspace}

\usepackage{color}
\definecolor{cyan}{rgb}{0,0.9,0.9}
\definecolor{orange}{rgb}{0.9,0.5,0}
\definecolor{magenta}{rgb}{1,0,1}
\definecolor{purple}{rgb}{0.8,0.4,0.8}
\definecolor{gray}{rgb}{0.5,0.5,0.5}
\definecolor{mygreen}{rgb}{0.1,0.8,0.1}
\definecolor{darkblue}{rgb}{0.0,0.0,0.6}

\begin{document}

\title{Relative binning for complete gravitational-wave parameter estimation with 
higher-order modes and precession, and applications to lensing and third-generation detectors}

\author{Harsh Narola$^{1,2}$}
\author{Justin Janquart$^{1,2}$}
\author{Quirijn Meijer$^{1, 2}$}
\author{K. Haris$^{1, 2, 3}$}
\author{Chris Van Den Broeck$^{1,2}$}

\affiliation{${}^1$Institute for Gravitational and Subatomic Physics (GRASP), 
Utrecht University, Princetonplein 1, 3584 CC Utrecht, the Netherlands}
\affiliation{${}^2$Nikhef -- National Institute for Subatomic Physics, 
Science Park 105, 1098 XG Amsterdam, the Netherlands}
\affiliation{${}^3$Department of Physics, National Institute of Technology, Kozhikode, Kerala 673601, India}

\date{\today}

\begin{abstract}
Once a gravitational wave signal is detected, the measurement of its source parameters is important 
to achieve various scientific goals. This is done through Bayesian inference, where the analysis cost 
increases with the model complexity and the signal duration. For typical binary black hole signals 
with precession and higher-order modes, one has 15 model parameters. With standard methods, such 
analyses require at least a few days. For strong gravitational wave lensing, where multiple images
of the same signal are produced, the joint 
analysis of two data streams requires 19 parameters, further increasing the complexity and run time. 
Moreover, for third generation detectors, due to the lowered minimum sensitive frequency, the signal 
duration increases, leading to even longer analysis times. With the increased detection rate, 
such analyses can then become intractable.
In this work, we present a fast and precise parameter estimation method relying on relative 
binning and capable of including higher-order modes and precession.
We also extend the method to perform joint Bayesian inference for lensed gravitational wave signals. 
Then, we compare its accuracy and speed to those of state-of-the-art parameter estimation routines 
by analyzing a set of simulated signals for the current and third generation of interferometers. 
Additionally, for the first time, we analyze some real events known to contain higher-order modes with relative 
binning. For binary black hole systems with a total mass larger than 
$50\, M_{\odot}$, our method 
is about 2.5 times faster than current techniques. This speed-up increases for lower masses,  
with the analysis time being reduced by a factor of 10 on average. In all cases, 
the recovered posterior probability distributions for the parameters match those 
found with traditional techniques.

\end{abstract}

\maketitle

\section{I\lowercase{ntroduction}}
\label{sec:intro} 
The Advanced Laser Interferometric Gravitational Wave Observatory (LIGO)~\cite{LIGOScientific:2014pky} 
and Advanced Virgo detector~\cite{VIRGO:2014yos} have reported 90 gravitational wave events across 
three observation runs combined~\cite{LIGOScientific:2021djp}. 
This number is expected to increase rapidly in the future due to the enhanced sensitivity of the 
existing detectors and the addition of new detectors to the network, such as KAGRA and 
LIGO-India~\cite{KAGRA:2018plz, LIGO-INDIA}. 
Furthermore, the third generation (3G) of gravitational wave detectors, 
such as Einstein Telescope (ET)~\cite{Punturo:2010zz, Maggiore2019ScienceTelescope, Branchesi:2023mws} 
and Cosmic Explorer (CE)~\cite{Reitze:2019iox}, is anticipated to be an order of magnitude more 
sensitive, leading to an even larger forecast number of detections. 

Once a GW signal is detected, an important next step is to measure its source properties. 
This is crucial for achieving numerous scientific objectives such as modeling the 
population of compact binary mergers~\cite{LIGOScientific:2021psn}, understanding the 
formation channels~\cite{Zevin:2020gbd, Rodriguez:2016kxx}, 
testing general relativity~\cite{LIGOScientific:2021sio}, 
probing cosmology~\cite{LIGOScientific:2021aug}, 
and measuring the neutron star equation of state~\cite{LIGOScientific:2018cki}. 
Since all these applications rely on parameter estimation, 
they would all benefit from improvements in analysis techniques. In particular, 
parameter estimation also plays a crucial role in searches for 
strongly-lensed gravitational wave signals, where the main idea is to look for repeated signals 
(``images'')  
with the same detector-frame parameters, only differing by an overall 
magnification, phase shift and time delay. Therefore, 
one analyzes the two events jointly, linking them through four 
lensing parameters~\cite{Janquart:2023osz, Lo:2021nae, Janquart:2021qov, Liu:2020par, Haris:2018vmn}. 
Since these analyses require a larger number of parameters, the computational run time is further 
increased. Hence this aspect of gravitational wave data analysis would also greatly 
benefit from improvements in parameter estimation methods. 

Nowadays, most of the state-of-the-art parameter estimation 
pipelines~\cite{Ashton2018Bilby:Astronomy, Veitch:2014wba, Biwer:2018osg, Wysocki:2019grj} rely on traditional Bayesian inference~\cite{Thrane:2018qnx} to measure the source 
parameters\footnote{We also note the emergence of novel techniques relying on machine learning and 
normalizing flows to perform rapid parameter inference~\cite{Green:2020hst, Dax:2021tsq, Dax:2022pxd, Langendorff:2022fzq}.}. 
However, even for signals observed by the LIGO and Virgo interferometers, which can be observable 
for a few seconds to minutes, the analyses can be computationally expensive as one needs to perform many likelihood evaluations. 
Traditionally, a single likelihood evaluation involves generating a simulated signal and calculating inner products with itself and the observed signal.  
Here, generating the waveform is the most expensive operation and its cost increases with the 
duration of the detected signal.
In the 3G detector era, where we will be able to observe binary signals for up to a day, 
the analysis times required by current methods would make it impossible to keep pace with  
the predicted detection rate~\cite{Samajdar:2021egv}.
Therefore, various methods to speed up parameter estimation have been 
proposed~\cite{Canizares:2014fya, Vinciguerra:2017ngf,  Morisaki:2020oqk, Morisaki:2021ngj, Cornish:2021lje, Smith:2021bqc, Islam:2022afg, Pathak:2022ktt, Fairhurst:2023idl, Morisaki:2023kuq}.

One such technique is relative 
binning~\cite{Zackay:2018qdy, Dai:2018dca, Leslie:2021ssu}, a method that accelerates 
the likelihood evaluations by using a reference waveform and a coarse frequency grid. 
For each iteration, one computes the proposal waveform generated by the samples only at 
certain grid points chosen so that the ratio between this waveform and the reference one can be 
approximated by a piece-wise linear function of frequency. These ratios can then be used to 
compute the inner products between the data and the waveform and the waveform with itself. 
The reduction in the number of points for which the waveform needs to be generated and the inner 
products computed at each iteration leads to a decrease in the run time. 
The faithfulness of the final result will depend mainly on two aspects: 
the quality of the reference parameters, and the choice in grid points, 
which needs to be such that the piece-wise linear approximation between the frequency nodes holds
to sufficient accuracy.

\begin{figure}
\includegraphics[width=\linewidth]{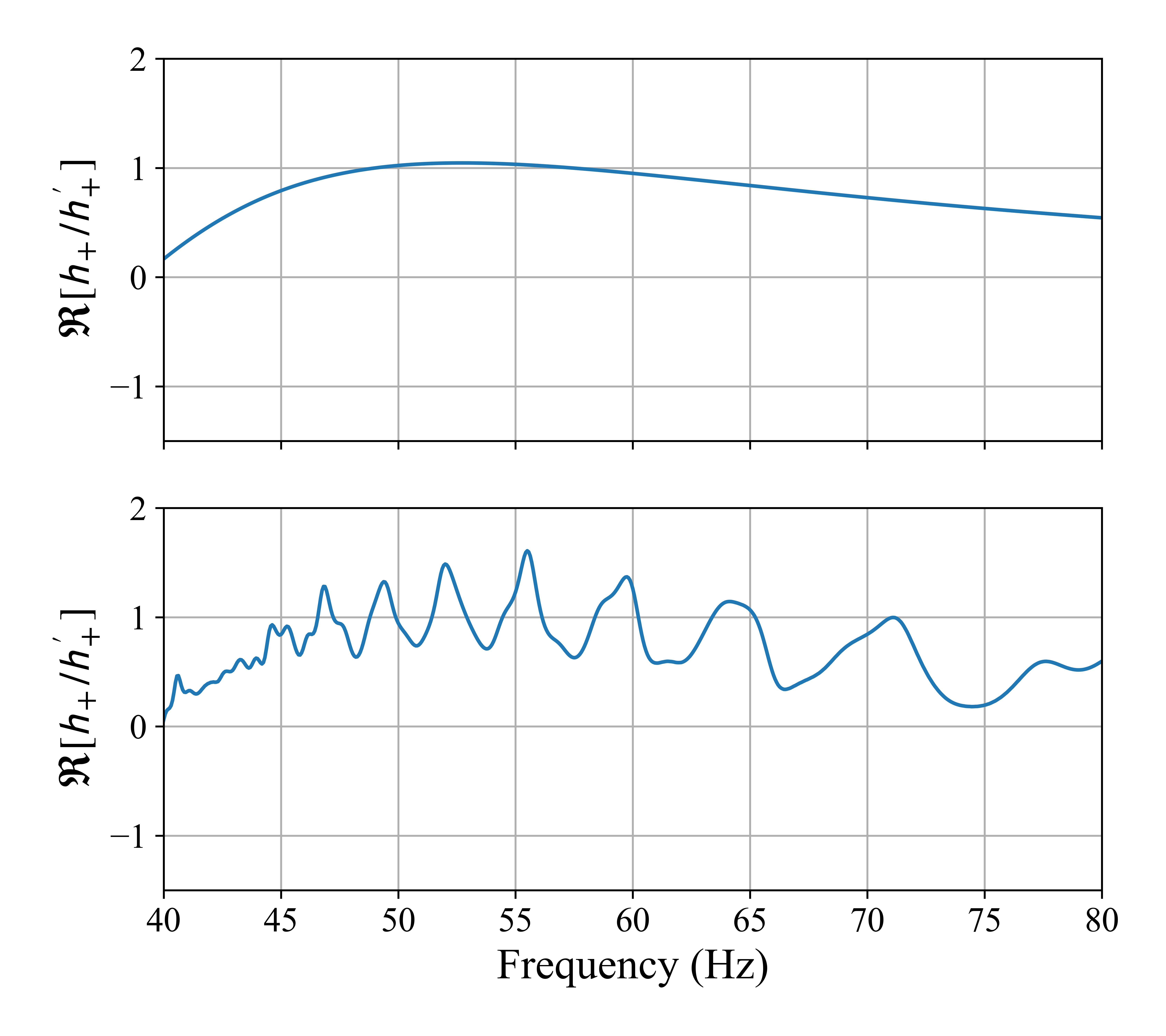}
\caption{Ratio (vertical axis) of the real part of the proposal to the fiducial waveform for 
the plus polarization as a function of frequency (horizontal axis). The top panel is for the 
IMRPhenomD waveform model with only the dominant mode and without precession, while the bottom panel is 
for the IMRPhenomXPHM waveform model containing HOMs and precession. 
The source parameters of the proposal and the fiducial waveform between the two plots are kept 
the same. The fiducial parameters are similar to the source parameters 
of GW190412~\cite{GW190412-detection}, a BBH merger with asymmetric masses and 
detectable HOM content, and the proposal parameters have a 5$\%$ shift from the fiducial 
chirp mass and mass ratio.}
\label{d-xphm-ratio}
\end{figure}

Each of the two aspects above leads to its own caveats. First, for a real event, the true 
parameters are unknown and one needs to find parameters that are close enough before 
starting the inference with relative binning. Some avenues are using the results 
from template-based searches~\cite{Usman:2015kfa, Sachdev:2019vvd, Aubin:2020goo} 
or running a maximum-likelihood estimator on the 
data~\cite{Srivastava:2018wvy, Storn:1997aa, 2020SciPy-NMeth, Finstad:2020sok} 
to obtain parameters representing the reference waveform. 
Of course, the quality of the fiducial waveform will depend on the way in which its 
parameters are chosen. For the choice in grid points, issues can arise when the 
complexity of the waveform model increases as smaller differences in parameters lead to 
larger oscillations in the waveform ratio~\cite{Leslie:2021ssu}. This is the case when one 
includes spin-orbit effects or higher-order modes 
(HOMs)~\cite{Hannam:2013oca, Pratten:2020ceb, Ossokine:2020kjp}. 
Therefore, the approximation works best when considering non-precessing waveforms 
with only the dominant mode, such as IMRPhenomD~\cite{Khan:2015jqa}. 
For more complete waveforms, such as IMRPhenomXPHM~\cite{Pratten:2020ceb} which 
incorporates both precession and HOMs, the ratio becomes more complicated, and the same 
difference between fiducial and proposal parameters for the waveform leads 
to a more complicated behavior for the waveform ratio, 
as illustrated in Fig.~\ref{d-xphm-ratio} where on sees larger 
oscillations for the IMRPhenomXPHM waveform. We also note that using the 
most complete waveforms for binary black hole (BBH) data analysis is crucial, 
since such effects can play important roles and their non-inclusion can bias the results~\cite{CalderonBustillo:2015lrt, Varma:2016dnf, Payne:2019wmy, Vijaykumar2022DetectionSignals, Janquart2021OnModes}.

Because of the larger oscillations in the ratio, analyzing data using HOMs and precession 
requires a much denser grid to keep a linear approximation between frequency points. 
Therefore, the resulting relative binning framework is not efficient. 
Recent work addresses this issue by decomposing the IMRPhenomXPHM waveform into its 
component modes and approximating the ratio for each mode individually~\cite{Leslie:2021ssu}. 
Even though their method produces accurate likelihood evaluations, the speed-up is not large enough 
to offer a major advantage over traditional parameter estimation techniques. 

In this work, we improve upon existing relative binning 
methods~\cite{Zackay:2018qdy, Dai:2018dca, Leslie:2021ssu} so that the technique can be used to perform 
parameter estimation with waveforms modeling HOMs and precession. 
This is accomplished by introducing new schemes to choose the coarse frequency grid and approximate the likelihood more effectively.
We optimize the implementation by using the just-in-time compiler \texttt{Numba}~\cite{lam2015numba}.  
Furthermore, we extensively test the speed and accuracy of our method by performing parameter inference on a set of 70 simulated BBH signals and a select set of observed BBH signals~\cite{LIGOScientific:2021djp}: GW190412~\cite{LIGOScientific:2020stg}, GW190814~\cite{LIGOScientific:2020zkf}, GW191103, and GW191105.
The set of simulated signals also includes a subset of events from the Gravitational Wave Transient 
Catalog (GWTC)~\cite{LIGOScientific:2021djp} as observed by a network of 3G detectors.  
Building upon the improvements made for parameter estimation of \textit{individual} events, we extend 
the method to perform \textit{joint} 
parameter estimation on images of strongly-lensed 
signals~\cite{Janquart:2023osz, Liu:2020par, Lo:2021nae, Janquart:2021qov}. 
We test the newly developed framework on a set of simulated lensed signals as well as the 
observed GW event pair GW191103--GW191105, which was investigated in the strong lensing 
searches due to its display of prototypical lensing signatures, even though the analyses
turned out to be inconclusive~\cite{LIGOScientific:2023bwz, Janquart:2023mvf}. 

Our paper is organized as follows. The key aspects of parameter estimation and the relative binning 
framework for individual events are explained in Sec.~\ref{sec:rb1}.
We summarize the main concepts of strong lensing, joint parameter estimation and the extension of the relative binning framework to perform joint parameter estimation in Sec.~\ref{sec:rb2}.  
We explain the algorithm to construct the coarse frequency grid in Sec.~\ref{sec:bin-selection}.
The results are presented in Sec.~\ref{sec:results}, where Sec.~\ref{subsec:indipe} contains the results of individual parameter estimation.
The latter is divided into two parts: Part~\ref{sss:hlv-sensitivity} shows results 
obtained for LIGO-Virgo interferometers, and Part~\ref{sss:3g-sensitivity} for 3G interferometers. 
Sec.~\ref{subsec:jpe} contains results of joint parameter estimation for lensed events, 
and Sec.~\ref{subsec:speed-up} details the speed-ups gained by the relative binning method for the different considered scenarios. 
Finally, we provide conclusions and discuss future directions in Sec.~\ref{sec:conclusions}.

\section{P\lowercase{arameter Estimation for Individual Events}}
\label{sec:rb1}

\begin{table}[htbp]
  \centering
  \begin{tabular}{| C{5cm} | C{3cm} |}
    \hline
    \textbf{Parameter} & \textbf{Population Generation Prior} \\
    \hline
    Chirp mass $(\mathcal{M}_c)$ &  Uniform($5M_{\odot},$ $100M_{\odot}$) \\
    \hline
    Mass ratio $(q)$ & Uniform(0.1, 10) \\
    \hline
   Dimensionless spin of the $i^{th}$ black hole $(a_i)$  & Uniform(0, 1) \\
    \hline
    Zenith angle between the spin and orbital angular momentum for 
    $i^{th}$ black hole $(\theta_i )$ &Sine($0, \pi$)\\
    \hline
     Difference between the azimuthal angles of the individual spin vector projections onto the orbital plane $(\Delta \phi)$ & Uniform($0, 2\pi$)\\
    \hline
    Difference between total and orbital angular momentum azimuthal angles $(\phi_{JL})$ &  Uniform($0, 2\pi$) \\
    \hline
    Right Ascension (RA) &  Uniform($0, 2\pi$)\\
    \hline
    Declination (DEC) &  Cosine($-\pi/2,$ $\pi/2$)\\
    \hline
   Angle between the line of sight and total angular momentum$ (\theta_{JN})$ &  Sine($0, \pi$)\\
    \hline
    Polarization angle $(\psi)$ &  Sine($0, \pi$)\\
    \hline
   Phase of coalescence $(\phi_c)$ &  Uniform($0, 2\pi$)\\
    \hline 
    Luminosity distance $(d_L)$ &  UniformSourceFrame(0.1 \text{Gpc}, 5 \text{Gpc}) \\
    \hline
    Time of coalescence $(t_c)$ &  Uniform(0s, 86400s) \\
    \hline
    Morse factor for image 1 $(n_1)$ &  Uniform\{0, 0.5, 1\} \\
    \hline
    Magnification of image 2 w.r.t image 1 $(\mu_{21})$ &  Uniform(0, 10) \\
    \hline
    Delay in arrival time between two images $(\Delta t_{21})$ &  Uniform(0s, 86400s) \\
    \hline
    Difference in Morse factor $(\Delta n_{21})$ &  Uniform\{0, 0.5, 1\} \\
    \hline
     \end{tabular}
    \caption{Parameters and corresponding priors used to simulate individual GW signals and the pairs of lensed GW signals. 
    For a detailed description of each prior probability distribution refer to~\cite{Ashton2018Bilby:Astronomy}.}
     \label{bbh_params}

\end{table}

The primary goal of parameter estimation is to calculate $p(\boldsymbol{\Theta}|\bm{d})$, the probability distribution 
of the source parameters $\boldsymbol{\Theta}$ -- ranging from $\mathcal{M}_c$ to $t_c$ in 
Table~\ref{bbh_params}-- given the gravitational wave data $\bm{d}$.

Using Bayes' theorem we have
\begin{equation}
p(\boldsymbol{\Theta}|\bm{d}) = \frac{\pi(\boldsymbol{\Theta})\,\mathcal{L}(\boldsymbol{\Theta})}{\mathcal{Z}} \,,
\end{equation}
where $\pi(\boldsymbol{\Theta}) = p(\boldsymbol{\Theta})$ is the prior probability distribution for $\boldsymbol{\Theta}$, 
$\mathcal{L}(\boldsymbol{\Theta}) = p(\bm{d}|\boldsymbol{\Theta})$ is the likelihood, and 
$\mathcal{Z} = \int  \pi(\boldsymbol{\Theta})\,\mathcal{L}(\boldsymbol{\Theta})\, \mathrm{d}\boldsymbol{\Theta}$ the evidence.

Approximating the noise in gravitational wave detectors as being Gaussian and stationary, the 
(log) likelihood can be expressed as \cite{Veitch:2009hd}
\begin{equation}
\label{eqn:indi-likelihood}
\ln \mathcal{L}(\boldsymbol{\Theta}) = -\frac{1}{2}(\bm{d}-\bm{h}(\boldsymbol{\Theta})|\bm{d}-\bm{h}(\boldsymbol{\Theta}))\,,
\end{equation}
where $\bm{h}(\boldsymbol{\Theta})$ represents the GW waveform as a function of the source parameters $\boldsymbol{\Theta}$ and $(\,.\,|\,.\,)$ 
denotes the noise-weighted inner product
\begin{equation}
(\bm{a}|\bm{b}) = \frac{4}{T}\Re\sum_f \frac{a(f)b^*(f)}{S_n(f)}.
\end{equation}
Here, $f$, $S_n(f)$, and $T$ respectively denote the frequency, one-sided noise power spectral density 
(PSD), and duration, while $\Re$ denotes the real part; the sum over frequencies will be discussed in detail below. 
Expanding $(\bm{d}-\bm{h}(\boldsymbol{\Theta})|\bm{d}-\bm{h}(\boldsymbol{\Theta}))$ in Eq.~\eqref{eqn:indi-likelihood}, we get

\begin{equation}
\label{eqn:logL}
\ln \mathcal{L}(\boldsymbol{\Theta}) = (\bm{d}|\bm{h}(\boldsymbol{\Theta})) - \frac{1}{2}(\bm{h}(\boldsymbol{\Theta})|\bm{h}(\boldsymbol{\Theta})) - \frac{1}{2}(\bm{d}|\bm{d}) \,.
\end{equation}

Likelihood-based inference involves calculating 
$(\bm{d}|\bm{h}(\boldsymbol{\Theta}))$ and $(\bm{h}(\boldsymbol{\Theta})|\bm{h}(\boldsymbol{\Theta}))$ 
for a large number of values for $\boldsymbol{\Theta}$ when it explores the parameter space. 
Therefore, we want to improve the speed in computing these quantities. 
Let us denote the term $(\bm{d}|\bm{h}(\boldsymbol{\Theta}))$ by $ \ln \mathcal{L}_{dh}$ and 
$(\bm{h}(\boldsymbol{\Theta})|\bm{h}(\boldsymbol{\Theta}))$ by $\ln \mathcal{L}_{hh}$ to rewrite Eq.~\eqref{eqn:logL} as 
\begin{equation}
\label{eqn:short-l}
\ln \mathcal{L}(\boldsymbol{\Theta}) = \ln \mathcal{L}_{dh} - \frac{1}{2} \ln \mathcal{L}_{hh} - \frac{1}{2}(\bm{d}|\bm{d}).
\end{equation}
Note that the last term does not depend on the source parameters and therefore need only be 
computed once for the data.

When performing parameter estimation, we are interested in knowing the shape of 
$\ln\mathcal{L}$ around its peak. In other words, we want to evaluate $\ln\mathcal{L}$ 
for waveforms that are {\it similar} to the waveform where $\ln\mathcal{L}$ peaks. 
If we choose a fiducial value of $\boldsymbol{\Theta}$ which is close enough to the peak, 
we can estimate $\ln\mathcal{L}(\boldsymbol{\Theta})$ by expanding it around the fiducial value. 
Let us denote the fiducial source parameters by $\boldsymbol{\Theta}'$ and the waveform at the 
fiducial parameters by $\bm{h}'$. For any proposal waveform $\bm{h}$, 
under the assumption that $\bm{h}$ and $\bm{h}'$ are close enough, we can find 
frequency bins $b =  [f_{min}, f_{max}]$ such that the ratio 
$\bm{h}/\bm{h}'$ can be linearly approximated as
\begin{equation}
\label{eqn:ratio}
\frac{h}{h'}(f) = r_1(b) + r_2(b)(f-f_c(b))+\mathcal{O}(f^2) 
\end{equation}
within a given bin, where $r_1(b)$ and $r_2(b)$ are expansion coefficients 
and $f_c$ denotes the central value of the frequency bin $b$. 

\begin{figure}
\includegraphics[width=\linewidth]{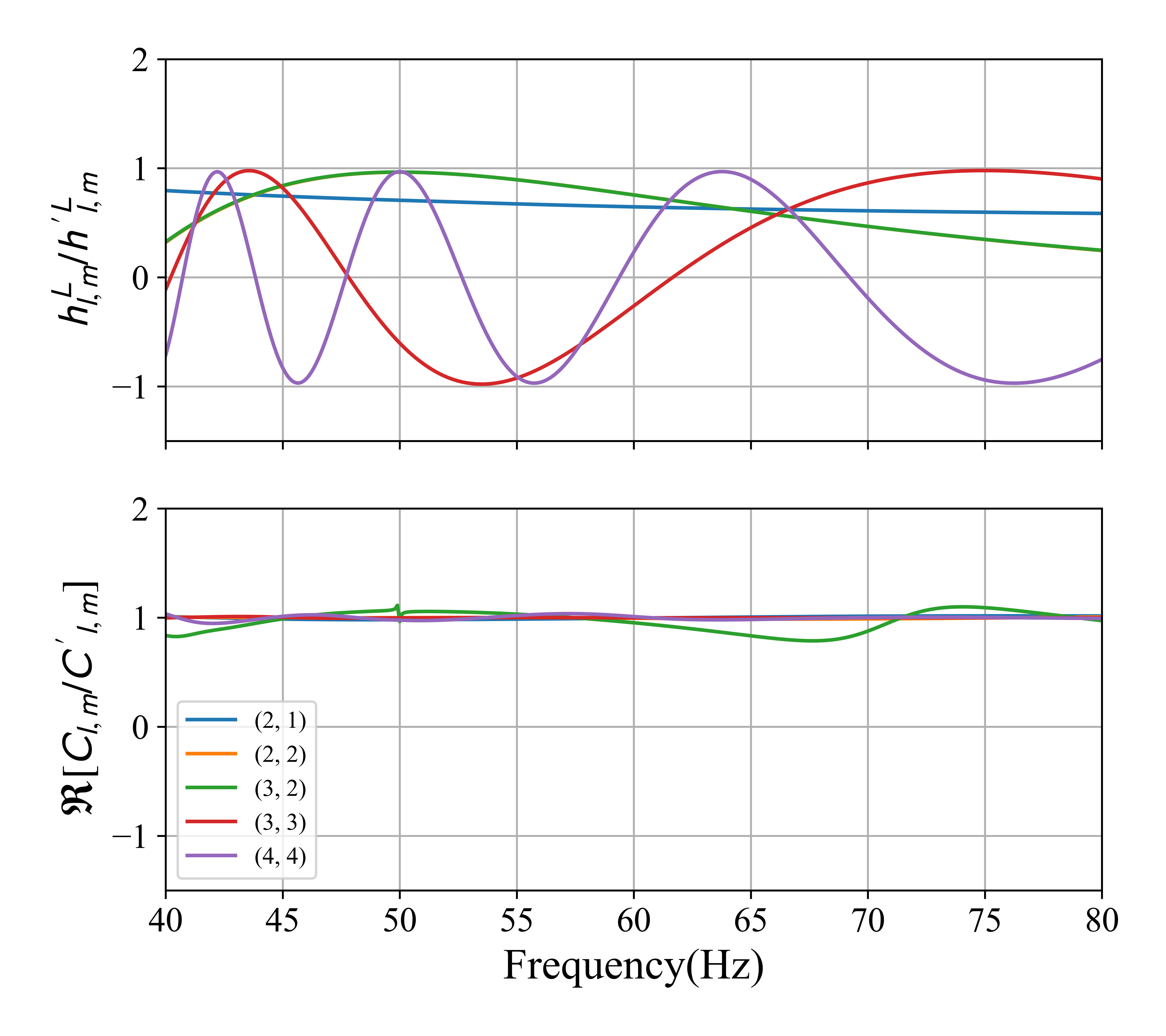}
\caption{The top (bottom) panel shows the ratio of $h^L$ (C) for the fiducial waveform to the proposal waveform for each mode as a function of frequency. 
The source and the fiducial parameters used to generate the plot are identical to the ones used in Fig.~\ref{d-xphm-ratio}.}
\label{hlc-plot}
\end{figure}

The approximation made in Eq.~(\ref{eqn:ratio}) performs better for the IMRPhenomD waveform compared to the IMRPhenomXPHM one because the ratio $\bm{h}/\bm{h}'$ varies more smoothly with $f$ for the former (see Fig.~\ref{d-xphm-ratio}).
Therefore, when using the IMRPhenomD waveform a few bins are sufficient to make the linear expansion from Eq.(\ref{eqn:ratio}) precise in each bin. This is not the case anymore for more complex waveform models. 
When using the IMPRhenomXPHM waveform, as the ratio is subject to more fluctuation, we need to consider a 
larger number of bins, which increases waveform generation costs and makes the likelihood 
evaluation slow. 

In order to make the approximation in Eq.~(\ref{eqn:ratio}) robust for precessing waveforms with HOMs, 
let us express the detector frame waveform for given $\boldsymbol{\Theta}$ into its 
component modes~\cite{Pratten:2020ceb} 
\begin{equation}
h(f) = \sum_{l, m}(F_+C^{+}_{l, m}(f)+F_\times C^{\times}_{l, m}(f))\,h^L_{l, m}(f)
\label{eqn:comp-mode}
\end{equation} 
Here, $F_+$ and $F_\times$ are the beam pattern functions of the interferometer, which depend 
only on extrinsic parameters; $\bm{h}^{L}$ denotes the waveform in the co-precessing frame 
(also known as the $L$-frame); and the indices $l$ and $m$ label the different modes. The coefficients $C^{+}_{l, m}$ and $C^{\times}_{l, m}$ account for 
the {\it twisting-up} procedure~\cite{Pratten:2020ceb} which transforms the waveform from a 
co-precessing frame to the inertial (or observer) frame. 
The benefit of handling each mode in the $L$-frame separately is that the ratio of a 
fiducial $\bm{h}'^{L}_{l, m}$ to a proposal $\bm{h}^{L}_{l, m}$ oscillates less compared to the 
ratio of the full waveforms $\bm{h}/\bm{h}'$. This can be seen by comparing the bottom panel of 
Fig.~\ref{d-xphm-ratio} with the top panel of Fig.~\ref{hlc-plot}. 
Therefore, a linear expansion similar to the one in Eq.~\eqref{eqn:ratio} 
can be made using the ratio $\bm{h}^{L}_{l, m}/\bm{h}'^L_{l, m}$ instead of 
$\bm{h}/\bm{h}'$ and will require a lower number of frequency bins for more complex models:
\begin{equation}
\label{eqn:l-ratio}
\frac{h^{L}_{l, m}}{h'^L_{l, m}} (f) = r_{1, l, m}(b) + r_{2, l, m} (b)(f-f_c(b)) + \mathcal{O}(f^2) \,,
\end{equation}

Now, let us denote the factor $(F_+C^{+}_{l, m}(f)+F_\times C^{\times}_{l, m}(f))$ in 
Eq.~(\ref{eqn:comp-mode}) by $\bm{C}_{l, m}$. We can perform similar expansions on the ratios 
$\bm{C}_{l, m}/\bm{C}'_{l, m}$:
\begin{equation}
\label{eqn:c-ratio}
\frac{C_{l, m}}{C'_{l, m}}(f) = s_{1, l, m}(b) + s_{2, l, m}(b)(f-f_c(b)) + \mathcal{O}(f^2).
\end{equation}
Below, the expansion coefficients will collectively be denoted 
$\mathcal{R} = \{\bm{r}_1, \bm{r}_2, \bm{s}_1, \bm{s}_2\}$. 
For now, we assume we can find frequency bins such that the piece-wise linear interpolations 
in Eqs.~(\ref{eqn:l-ratio}) and (\ref{eqn:c-ratio}) are 
valid. How such bins can be found will be explained in Sec.~\ref{sec:bin-selection}.

Once the frequency bins are made based on the chosen fiducial waveforms, we can pre-compute some 
\textit{summary data}~\cite{Leslie:2021ssu} based only on the reference waveform and useful for the 
computation of the inner products in Eq.~(\ref{eqn:short-l}). These are:
\begin{eqnarray}
&& W_{l, m}(b) = \frac{4}{T} \sum_{f\in b}\frac{d(f)h'^{L*}_{l, m}(f)C'^{*}_{l, m}(f)}{S_n(f)} \, , 
\nonumber \\
&& X_{l, m}(b) = \frac{4}{T} \sum_{f\in b}\frac{d(f)h'^{L*}_{l, m}(f)C'^*_{l, m}(f)}{S_n(f)}(f-f_c(b)) \,, 
\nonumber\\
&& Y_{l, m, \bar{l}, \bar{m}}(b) = \frac{4}{T} \sum_{f\in b}\frac{h'^L_{l, m}(f)C'_{l, m}(f)h'^{L*}_{\bar{l}, \bar{m}}(f)C'^*_{\bar{l}, \bar{m}}(f)}{S_n(f)} \,, 
\nonumber\\
&&Z_{l, m, \bar{l}, \bar{m}}(b)  \nonumber\\
&&= \frac{4}{T} \sum_{f\in b}\frac{h'^L_{l, m}(f)C'_{l, m}(f)h'^{L*}_{\bar{l}, \bar{m}}(f)C'^*_{\bar{l}, \bar{m}}(f)}{S_n(f)}(f-f_c(b))\,.
\nonumber\\
\end{eqnarray}

Note that the summary data 
$\mathcal{D} = \{\bm{W}_{l, m}, \bm{X}_{l, m}, \bm{Y}_{l, m, \bar{l}, \bar{m}}, \bm{Z}_{l, m, \bar{l}, \bar{m}} \}$ 
need to be computed only once per analysis and they depend only on the choice of the fiducial 
parameters $\boldsymbol{\Theta}'$. On the other hand, the coefficients $\mathcal{R}$ need to be computed for every 
proposal waveform as they depend on $\boldsymbol{\Theta}$ and $\boldsymbol{\Theta}'$.  

Using the summary data $\mathcal{D}$ and the coefficients $\mathcal{R}$, we can estimate $\ln\mathcal{L}_{dh}$ and $\ln\mathcal{L}_{hh}$ in Eq.~\eqref{eqn:short-l} 
for given $\boldsymbol{\Theta}$ as
\begin{widetext}
\begin{eqnarray}
\label{eqn:dh}
\ln\mathcal{L}_{dh} &=& \Re \sum_{(l, m)} \sum_{b}  \left\{  W_{l, m}(b)\left[r_{1, l, m}^*(b) s_{1, l, m}^*(b)\right] +X_{l, m}(b)\left[r_{1, l, m}^*(b)s_{2, l, m}^*(b) + r_{2, l, m}^*(b) s_{1, l, m}^*(b) \right]  \right\} \,, \\
\label{eqn:hh}
\ln\mathcal{L}_{hh} &=&\Re  \sum_{(l, m), (\bar{l}, \bar{m})} \sum_{b} \left\{  Y_{l, m, \bar{l}, \bar{m}}(b) \left[ r_{1, l, m}(b)s_{1, l, m}(b) r^*_{1, \bar{l}, \bar{m}}(b) s^*_{1, \bar{l}, \bar{m}}(b) \right] \right. \nonumber \\
&& \qquad\qquad\qquad\quad +  \left. Z_{l, m, \bar{l}, \bar{m}}(b) \left[r_{1, l, m}(b) r_{1, \bar{l}, \bar{m}}^*(b) \left( s_{1, \bar{l}, \bar{m}}^*(b) s_{2, l, m}(b) + s_{1, l, m}(b) s_{2, \bar{l}, \bar{m}}^*(b) \right) \right. \right. \nonumber \\ 
&& \qquad\qquad\qquad\qquad\qquad\qquad\qquad  + \left. \left.  s_{1, l, m}(b)s_{1, \bar{l}, \bar{m}}^*(b)\left(r_{1, \bar{l}, \bar{m}}^*(b) r_{2, l, m}(b) + r_{1, l, m}(b) r_{2, \bar{l}, \bar{m}}^*(b)\right) \right] \right\},
\end{eqnarray}
\end{widetext}
which together with Eq.~(\ref{eqn:short-l}) enables us to approximate the likelihood. Provided the number of bins is reduced compared to the usual method, this approach should offer a more rapid likelihood evaluation, leading to a more rapid parameter inference.

\section{P\lowercase{arameter Estimation for Strong Lensing}}
\label{sec:rb2}
When analyzing a strongly-lensed gravitational wave signal that has been split into several images, one can perform joint parameter estimation, i.e.~one can analyze the images at the same time 
taking into account the common parameters in addition to the ones introduced by lensing~\cite{Janquart:2023osz, Liu:2020par, Lo:2021nae, Janquart:2021qov}. 
For two observable images, the joint analysis requires 19 parameters (see Table~\ref{bbh_params}), hence 4 more than single unlensed signal parameter 
estimation. If more than two images are present, each additional image will introduce three more parameters, thereby further increasing the computational time.
In this section, we recall the main aspects of strong lensing and describe the extension of the relative binning method to perform joint parameter estimation. For 
simplicity we focus on the case of two images, but the extension to $N > 2$ images is straightforward.

\subsection{Effect of strong lensing on gravitational-wave signals}

When a GW signal is strongly lensed, it is split into multiple images, each undergoing a (de-)magnification, time delay and overall phase shift~\cite{Dai2020SearchO2}. 
The $j^{\mathrm{th}}$ lensed image ($\bm{h}^l_j$) is related to the unlensed waveform ($\bm{h}^u$) through 
\begin{equation}
h^l_j (f; \boldsymbol{\Theta}, \mu_j, t_j, n_j) = \sqrt{\mu_j} h^{u}(f;\boldsymbol{\Theta})e^{2i\pi f t_j -in_j\pi} \,,
\end{equation}
where $\mu_j$, $t_j$ and $n_j$ denote the magnification factor, arrival time, and Morse factor for the $j^{\text{th}}$ image, respectively. We refer to them as the \textit{lensing parameters} of the $j^{\text{th}}$ image.
 
For a given image, these lensing parameters can not all be separately measured. Indeed, the magnification and the time delay are degenerate with the luminosity distance and the time of coalescence, 
respectively~\cite{Dai:2017huk, Lo:2021nae, Janquart:2021qov}. Therefore, we can absorb $\sqrt{\mu_j}$ into the luminosity distance $D_L$, and $t_j$ into the coalescence time $t_c$ 
to define the following effective quantities for the $j^{th}$ image:
\begin{align}
D_{L, j}^{\text{eff}} &= \frac{D_L}{\sqrt{\mu_j}} \,, \\
t_{c, j}^{\text{eff}} &= t_c+t_j \,.
\end{align}
These quantities can be measured from the GW data. 
On the other hand, the Morse phase is not necessarily degenerate with the phase of coalescence, as the presence of HOMs can break the possible phase-Morse factor 
degeneracy~\cite{Dai:2017huk, Ezquiaga2020PhaseWaves, Janquart:2021nus}

As mentioned above, for simplicity we focus on the scenario where two strongly-lensed images are analyzed. We refer to the image that arrives first at the detector as image 1, 
and denote its observed parameters by $\boldsymbol{\Theta}_1$\footnote{Note that in the context of strong lensing $\boldsymbol{\Theta}_1$ includes the Morse factor, and the effective luminosity and time delay for the first image.}. Since some of the lensing parameters are not measurable on their own, let us define \emph{relative lensing parameters} linking the lensing parameters of the two images:
\begin{align}
\label{eqn:mu21}
\mu_{21} = \mu_2/\mu_1  &= \left(D_{L, 1}^{\text{eff}} / D_{L, 2}^{\text{eff}}\right)^2  \\\
\Delta t_{21} &= t^{\text{eff}}_{c, 2}-t^{\text{eff}}_{c, 1} \\
\label{eqn:n21}
\Delta n_{21} &= n_2 - n_1 
\end{align}
We collectively denote the relative lensing parameters by $\boldsymbol{\Phi}_{21} = \{\mu_{21}, \Delta t_{21}, \Delta n_{21}\}$. 
Table \ref{tab:notation-table} summarizes all the parameters defined in the context of strong lensing. 
The relative lensing parameters can be used to express the GW from the second image in terms of the waveform for the first image:
\begin{equation}
\label{eqn:wf-mapping}
h^l_2(f; \boldsymbol{\Theta}_2) = \sqrt{\mu_{21}}h^{l}_1(f; \boldsymbol{\Theta}_1)e^{2i\pi f \Delta t_{21} - i\Delta n_{21}\pi}.
\end{equation}

\begin{table}
\begin{center}
\begin{tabular}{| c | p{1 in} | p{1.7 in} |}
\hline 
 Notation & Description & Parameters  \\ \hline 
 $\boldsymbol{\Lambda}$ & Source parameters common between the two images & \{$\mathcal{M}_c, \eta, a_1, a_2, \theta_1, \theta_2, \Delta \phi, \phi_{JL}$, $\text{RA}, \text{DEC}, \theta_{JN}, \psi, \phi \}$ \\ \hline
$\boldsymbol{\Theta}_1$ & Observed parameters of image 1 & $ \boldsymbol{\Lambda} \cup \{D^{\text{eff}}_{L, 1}, t^{\text{eff}}_{c, 1}, n_1\}  $ \\ \hline 
$\boldsymbol{\Theta}_2$ & Observed parameters of image 2 & $\boldsymbol{\Lambda} \cup \{D^{\text{eff}}_{L, 2}, t^{\text{eff}}_{c, 2}, n_2 $\} \\ \hline 
$\boldsymbol{\Phi}_{21}$ & Lensing parameters of image 2 relative to image 1 & ${\mu_{21}, \Delta t_{21}, \Delta n_{21}}$ \\ \hline 
\end{tabular}
\end{center}
\caption{Lensing parameters and source parameters defined in context of a pair of strongly lensed GW signals.}
\label{tab:notation-table}
\end{table}

\subsection{Joint parameter estimation for strong lensing}

Under the hypothesis that two GW events are strongly lensed, the two data streams can be written as $\bm{d}_1$ and $\bm{d}_2$, and the joint parameter estimation aims to measure 
\begin{equation}
p(\boldsymbol{\Theta}_1, \boldsymbol{\Theta}_2 | \bm{d}_1, \bm{d}_2, \mathcal{H}_l)\, ,
\end{equation}
where $\mathcal{H}_l$ denotes the strong lensing hypothesis. 
However, given $\boldsymbol{\Theta}_1$ and $\Phi_{21}$, we can calculate $\boldsymbol{\Theta}_2$ using Eq.(\ref{eqn:mu21})-(\ref{eqn:n21}). 
Therefore, we can equivalently infer
\begin{equation}
p(\boldsymbol{\Theta}_1, \Phi_{21} | \bm{d}_1, \bm{d}_2, \mathcal{H}_l)\, .
\end{equation}
Using Bayes' theorem we can write 
\begin{eqnarray}
\label{eqn:jpe-eqn}
&& p(\boldsymbol{\Theta}_1, \boldsymbol{\Phi}_{21} | \bm{d}_1, \bm{d}_2, \mathcal{H}_l) \nonumber\\
&& = \frac{\pi(\boldsymbol{\Theta}_{1}, \boldsymbol{\Phi}_{21} | \mathcal{H}_l) p( \bm{d}_1, \bm{d}_2 | \boldsymbol{\Theta}_1, \boldsymbol{\Phi}_{21}, \mathcal{H}_l)}{\mathcal{Z}_l} \,,
\end{eqnarray}
where $\pi(\boldsymbol{\Theta}_1, \boldsymbol{\Phi}_{21} | \mathcal{H}_l)$ is the prior on $\boldsymbol{\Theta}_1$ and $\boldsymbol{\Phi}_{21}$, 
$p( \bm{d}_1, \bm{d}_2 | \boldsymbol{\Theta}_1, \boldsymbol{\Phi}_{21}, \mathcal{H}_l)$ is the likelihood, and $\mathcal{Z}_l$ is the evidence under the strong lensing 
hypothesis\footnote{Excluding selection and population effects; see Ref.~\cite{Lo:2021nae} for more details.}: 
\begin{equation}
\mathcal{Z}_l = \int \mathrm{d}\boldsymbol{\Theta}_1 \mathrm{d}\boldsymbol{\Phi}_{21} \,  p(\boldsymbol{\Theta}_1, \boldsymbol{\Phi}_{21} | \mathcal{H}_l) \,  p(\bm{d}_1, \bm{d}_2 | \boldsymbol{\Theta}_1, \boldsymbol{\Phi}_{21}, \mathcal{H}_l) \,. 
\label{eqn:zL}
\end{equation}

Additionally, the joint likelihood in Eq.(\ref{eqn:jpe-eqn}) can be decomposed as
\begin{align}
\label{eqn:jpe-small-eqn}
p(\bm{d}_1, \bm{d}_2 | \boldsymbol{\Theta}_1, \boldsymbol{\Phi}_{21}) &= p(\bm{d}_1|\boldsymbol{\Theta}_1, \boldsymbol{\Phi}_{21}) \, p(\bm{d}_2 | \boldsymbol{\Theta}_1, \boldsymbol{\Phi}_{21}) \nonumber \\ 
& = p(\bm{d}_1 | \boldsymbol{\Theta}_1) \,  p(\bm{d}_2 | \boldsymbol{\Theta}_1, \boldsymbol{\Phi}_{21}) \,,
\end{align}
hinting at a possible way to perform relative binning analyses, where each of the terms above can be computed using the relative binning approximation before being combined.

\subsection{Relative binning for strong lensing}

The likelihood $p(\bm{d}_1| \boldsymbol{\Theta}_1)$ in Eq.~\eqref{eqn:jpe-small-eqn} is similar to the one expressed in Eq.(\ref{eqn:indi-likelihood}), except for the Morse factor, 
which is a simple constant phase shift to add to the waveform. So, the framework in Eqs.~(\ref{eqn:l-ratio})-(\ref{eqn:hh}) of Sec.~\ref{sec:rb1} to estimate the likelihood 
for individual events is also applicable here.
First, we choose appropriate fiducial parameters $\boldsymbol{\Theta}'_1$ and calculate the summary data $\mathcal{D}_1$ and the coefficients $\mathcal{R}_1$ for event 1.
Using $\mathcal{D}_1$ and $\mathcal{R}_1$, we can compute $\ln\mathcal{L}_{d_1h_1}$ and $\ln\mathcal{L}_{h_1h_1}$ as shown in Eqs.~(\ref{eqn:dh}) and 
(\ref{eqn:hh}). 
The likelihood $p(\bm{d}_1|\boldsymbol{\Theta}_1)$ is then obtained by
\begin{equation}
\label{eqn:ll-one}
\ln p(\bm{d}_1|\boldsymbol{\Theta}_1) = \ln\mathcal{L}_{d_1h_1} - \frac{1}{2} \ln\mathcal{L}_{h_1h_1} - \frac{1}{2} (\bm{d}_1 | \bm{d}_1) \,,
\end{equation}
where $(\bm{d}_1 | \bm{d}_1)$ again needs to be computed only once for the given GW data.  

The waveform $\bm{h}_1$ used to calculate the likelihood of $\bm{d}_1$ given $\boldsymbol{\Theta}_1$ can be rescaled 
to obtain $\bm{h}_2$ using Eq.(\ref{eqn:wf-mapping}) and $\boldsymbol{\Phi}_{21}$. 
The rescaled waveforms $\bm{h}_2$ can be used when sampling the likelihood of $\bm{d}_2$ given $\boldsymbol{\Theta}_2$;  
this approach saves additional computational costs that would be incurred in generating completely new waveforms. 
The rest of the steps are similar to the ones performed when calculating $p(\bm{d}_1|\boldsymbol{\Theta}_1)$, and we obtain 
\begin{equation}
\label{eqn:ll-two}
\ln p(\bm{d}_2|\boldsymbol{\Theta}_1, \boldsymbol{\Phi}_{21}) = \ln\mathcal{L}_{d_2h_2}-\frac{1}{2} \ln\mathcal{L}_{h_2h_2} - \frac{1}{2} (\bm{d}_2 | \bm{d}_2) \,.
\end{equation}

The joint likelihood defined in Eq.(\ref{eqn:jpe-small-eqn}) can be then calculated from Eqs.~(\ref{eqn:ll-one}) and (\ref{eqn:ll-two}) as
\begin{eqnarray}
&& \ln p(\bm{d}_1, \bm{d}_2 | \boldsymbol{\Theta}_1, \boldsymbol{\Phi}_{21}) \nonumber\\
&& = \ln\mathcal{L}_{d_1h_1} - \frac{1}{2} \ln\mathcal{L}_{h_1h_1} - \frac{1}{2} (\bm{d}_1 | \bm{d}_1) \nonumber \\
&& \quad + \ln\mathcal{L}_{d_2h_2}-\frac{1}{2} \ln\mathcal{L}_{h_2h_2} - \frac{1}{2} (\bm{d}_2 | \bm{d}_2) \,.
\end{eqnarray}
The combination of the relative binning for the first image and the waveform recycling to produce waveforms for the second image leads 
to a faster joint parameter estimation for strongly-lensed GW signals.

\section{B\lowercase{in Selection}}
\label{sec:bin-selection}

The selection of bins is an important step when using the relative binning method, as it determines not only the quality 
of the approximation but also the final speed-up obtained. 
The objective of this step is to identify a minimal set of frequency bins in which the ratios of proposal waveform components 
($\boldsymbol{h}^L_{l, m}$ and $\boldsymbol{C}_{l, m}$) to fiducial waveform components are linear to sufficient approximation.
We refer to a set of frequency bins which satisfy this condition as an RBGrid. 
The RBGrid is used to compute the summary data $\mathcal{D}$, and, once the sampling begins, 
to compute coefficients $\mathcal{R}$ for the points in parameter space proposed by the sampler. 

To initialize the bin selection algorithm, we need the following quantities: 
a uniform frequency grid with spacing $\Delta f = 1/T$, where the time $T$ is sufficiently long so as to easily accommodate the length of the signal; 
a fixed value of ``total error'' on log likelihood, $\epsilon$ (whose role will become apparent below); 
an initial value $N_0$ for the total number of frequency bins;  
a set of fiducial waveform parameters; and a set of test parameters. For definiteness we set  
$\epsilon = 0.01$ and $N_0 = 200$. In simulations, the fiducial parameters are set to the injection parameters, while for real events they are set to the 
maximum likelihood parameters obtained from earlier analyses, also as explained below. To choose a set of test parameters, we perturb the chirp mass and the 
mass ratio of the fiducial parameters by a random relative value in a $[ -10\%, 10\%]$ interval, and the rest of the parameters are kept the same. 

Once we have everything described in the paragraph above, we can construct the RBGrid as follows. 
We start from the first bin of the uniform frequency grid. 
We merge the first bin with the next one and calculate the partial relative binning log likelihoods $\ln\mathcal{L}_{dh}$ and $\ln\mathcal{L}_{hh}$ 
defined in Eqs.~\eqref{eqn:dh}-\eqref{eqn:hh}, 
using as test quantities the edges of the combined frequency bin, and the fiducial waveform parameters.
We also compute the corresponding exact partial likelihoods obtained by using all frequencies within the bins and not just the edges. 
If the absolute difference between the partial relative binning likelihoods and the partial exact likelihoods is less than $\epsilon/\sqrt{N_0}$, 
we add a subsequent bin into it and repeat. 
If the difference is larger than that, we do not add any more bins and move to the next bin and repeat. 
This continues until the maximum frequency of the uniform grid is reached. The value of $N_0$ is then updated to the total number of bins obtained 
through the above process, and the same procedure is started again with the same $\epsilon$ but the updated $N_0$. 
The algorithm terminates when $N_0$ ceases to change, and the edges of the merged bins in the final iteration become the RBGrid. 

The choice of $\epsilon$ controls the trade-off between speed and accuracy, where a larger value increases speed but reduces accuracy. By contrast, we have verified
that the bin selection procedure is not very sensitive to the initial value of $N_0$; it mainly affects the convergence rate of the algorithm. 
Setting $N_0$ equal to the length of the uniform grid would eliminate the need for this additional tuning parameter, but this would slow down the bin selection
algorithm as a whole.

The construction of the RBGrid when performing joint parameter estimation on pairs of lensed images is straightforward.
Besides the fiducial parameters, test parameters, and threshold value $\epsilon$, the RBGrid also depends on the signal-to-noise ratio (SNR) and PSD of each event. 
As the latter two quantities and the lensing parameters can vary significantly between two events in a pair,  
we construct the RBGrid for each event separately and use each of them to calculate $\{\mathcal{R}_1, \mathcal{D}_1\}$ and $\{\mathcal{R}_2, \mathcal{D}_2\}$ during the 
data analysis. 

Our implementation of the bin selection and relative binning parameter estimation framework can be found in Ref~\cite{janquart:2022}.   

\section{R\lowercase{esults}}
\label{sec:results}
To evaluate the performance of the relative binning framework presented in the previous sections, we conduct parameter estimation experiments 
on a large set of simulated GW signals and a select set of observed GW signals from the GWTC-3 catalogue~\cite{LIGOScientific:2021djp}. 
The goal is to compare the accuracy and speed of the parameter estimation runs performed with and without relative binning. 
Henceforth we refer to the method without relative binning as the ``regular'' method. 
Our results are divided over three subsections. 
The first two subsections focus on comparing the accuracy of the relative binning method and the regular one, 
while the third subsection address the speed-up achieved by the relative binning method.

In what follows, we use the \textsc{Dynesty} (version 2.1.1)~\cite{Speagle:2019ivv} sampler and $\textsc{Bilby}$ (version 2.1.0)~\cite{Ashton:2018jfp} for regular parameter estimation. 
Our relative binning framework is also implemented in such a way that it is compatible with \textsc{Bilby}. 
When we need to perform joint parameter estimation runs on lensed images with the regular method, we rely on \textsc{Golum}~\cite{Janquart:2023osz}.  
The starting frequency is equal to 20 Hz for the LIGO and Virgo interferometers and 5 Hz for the 3G interferometers.
For the latter, we consider a triangular ET~\cite{Punturo:2010zz, Maggiore2019ScienceTelescope, Branchesi:2023mws} with an arm length of 10 km and one CE~\cite{Reitze:2019iox} with an arm length of 40 km. 
The duration of the simulated signals is estimated using the \texttt{XLALSimInspiralTaylorF2ReducedSpinChirpTime}~\cite{lalsuite-ref} routine with additional padding of 4 seconds. 
The sampling frequency is equal to 2048 Hz if the duration is shorter than 5 seconds and 4096 Hz otherwise. 
Simulated signals are injected into Gaussian noise generated from a reference PSD chosen for each detector.
For the LIGO and the Virgo interferometers, the reference PSDs were taken from~\cite{KAGRA:2013rdx} and for 3G interferometers, from~\cite{threeGdesign}.
All of the configuration parameters are read from the GWOSC\footnote{Gravitational Wave Open Science Center (https://gwosc.org/eventapi/html/GWTC/)} data release~\cite{LIGOScientific:GWTC3DR, LIGOScientific:GWTC2.1DR} when analyzing real GW signals. 
For the simulated events, the fiducial parameters are set equal to the injected parameters, while for the observed events, they are set to the maximum likelihood parameters recovered when analyzing the data with the IMRPhenomXPHM waveform~\cite{LIGOScientific:GWTC3DR, LIGOScientific:GWTC2.1DR}.
The test parameters are derived by perturbing the chirp mass and mass ratio of the fiducial parameters with a random value taken from a $[-10\%, 10\%]$ interval. 
For $\mathcal{M}_c$, the prior used for parameter inference is uniform and centered at the fiducial value with a width of 5 $M_\odot$ on each side. 
For $t_c$ and $\Delta t_{21}$, the prior is also uniform and centered at the fiducial value with a width of $0.1$ seconds on each side. 
For $n_{1}$ and $\Delta n_{21}$, prior is uniform and continuous over the intervals [0, 1] and [0, 1.5], respectively.  
For all other parameters, the prior is the same as the population generation prior mentioned in Table~\ref{bbh_params}. 
We use the IMRPhenomXPHM waveform model for the analyses, and we include the following modes: (2, 1), (2, 2), (3, 2), (3, 3), (4, 4).  
Finally, we note that we do not include calibration priors in our analyses of observed 
events, but these are not expected to strongly modify the posterior distributions~\cite{Sun:2020wke, Sun:2021qcg}.

\subsection{Individual \lowercase{Parameter Estimation}}
\label{subsec:indipe}

In this subsection, we present the results for the individual parameter estimation obtained using our relative binning framework. We divided this subsection into two parts: 
results related to the LIGO-Virgo interferometers are given in \ref{sss:hlv-sensitivity}, while \ref{sss:3g-sensitivity} contains the results for the 3G network of interferometers. 

\begin{figure}
\includegraphics[width=\linewidth]{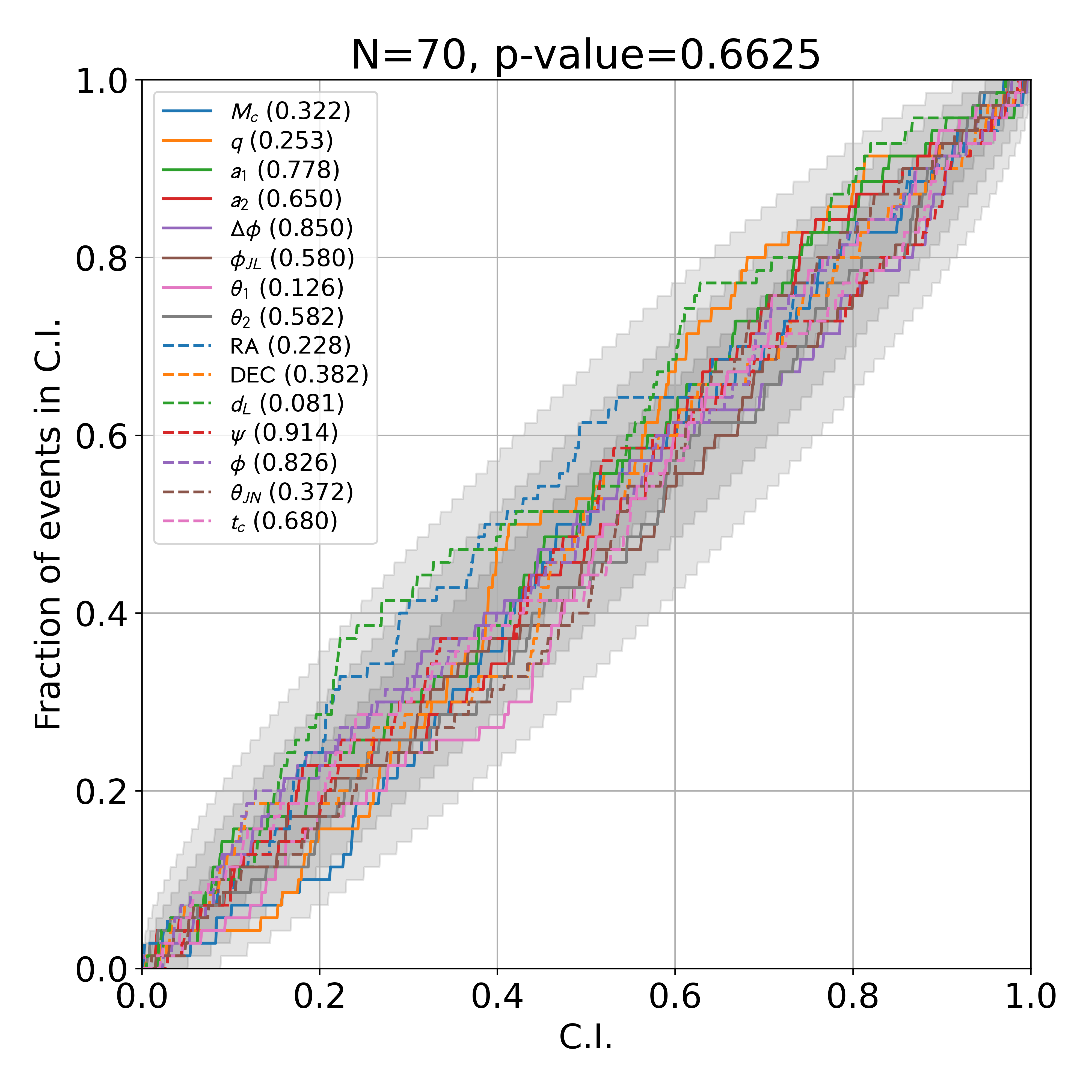}
\caption{P-P plot showing the results of parameter estimation performed on individual events using relative binning. A total of 70 simulated BBH signals were analyzed.
The numbers in the brackets of the legend show the p-values of the KS tests. The combined p-value of all parameters is 0.6625, consistent with the hypothesis 
that individual p-values were derived from a uniform distribution as expected. The shaded regions show $1\sigma, 2\sigma, 3\sigma$ confidence intervals in 
decreasing order of opacity.}
\label{pp-plot}
\end{figure}

\begin{figure}
\includegraphics[width=\linewidth]{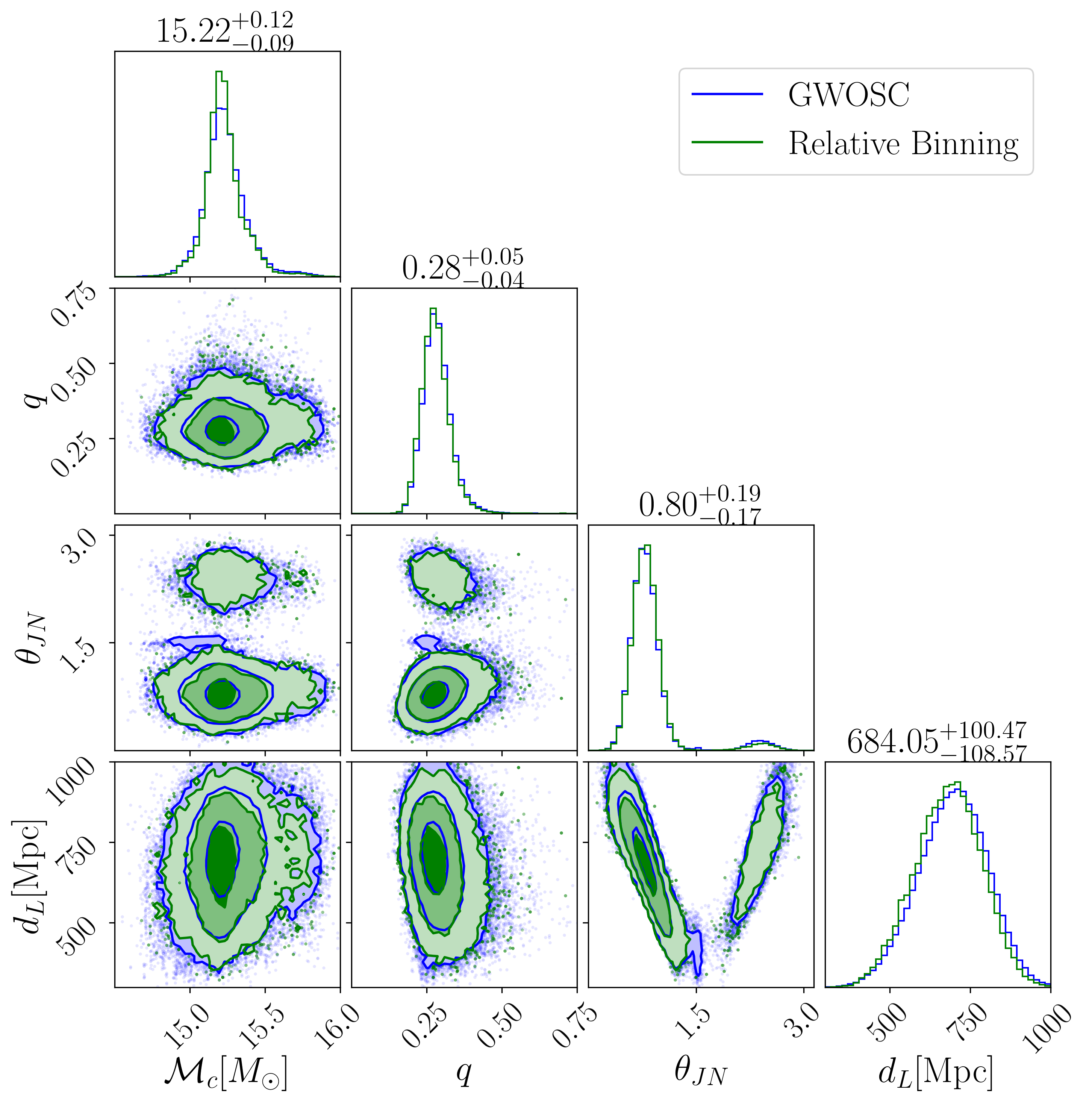}
\caption{Comparison of the posterior probability distributions obtained using relative binning (green curves) and the GWOSC data release (blue curves) for GW190412.}
\label{gw190412}
\end{figure}

\begin{figure}
\includegraphics[width=\linewidth]{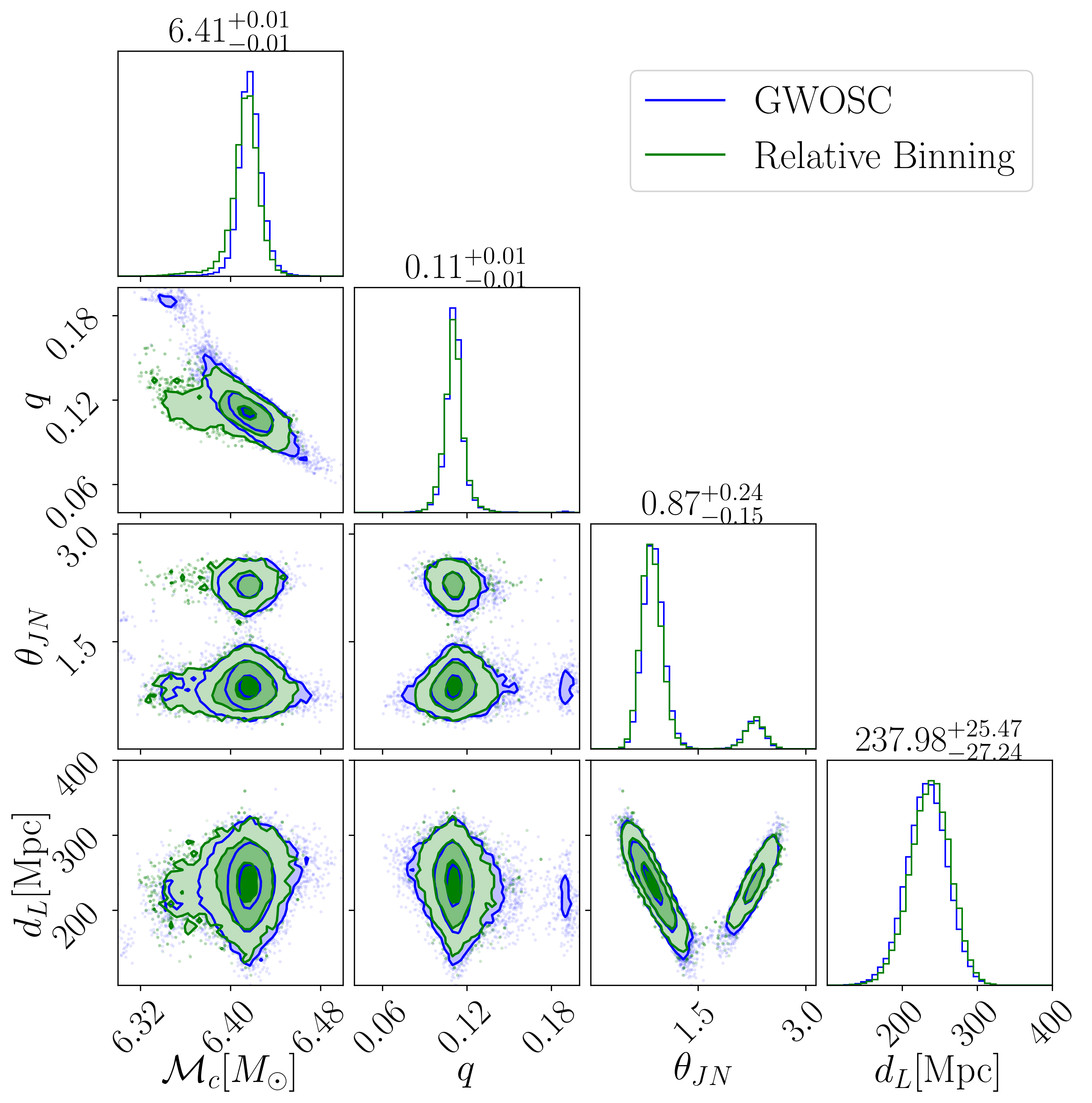}
\caption{Comparison of the posterior probability distributions obtained using relative binning (green curves) and the GWOSC data release (blue curves) for GW190814.}
\label{gw190814}
\end{figure}

\subsubsection{LIGO-Virgo \lowercase{Interferometers}}
\label{sss:hlv-sensitivity}

For a parameter estimation method to be trustworthy, it should recover the injected parameters within a given confidence interval for a corresponding fraction of injections; the 
extent to which it does so can be assessed using a so-called P-P plot~\cite{Sidery:2013zua, Veitch:2014wba, Berry:2014jja, Singer:2015ema, Romero-Shaw:2020owr}. 
To this end we simulate a population of 70 BBH signals by sampling the appropriate source parameters from the prior distributions specified in Table~\ref{bbh_params}. 
We adjust the luminosity distance in such a way that each event has a network SNR above 13. 
The network consists of the two LIGO~\cite{LIGOScientific:2014pky} and the Virgo interferometers~\cite{VIRGO:2014yos} at their design sensitivity~\cite{KAGRA:2013rdx}. 
We perform parameter estimation on each event and show the collective results in Fig.~\ref{pp-plot}, representing a P-P plot for our relative binning framework. Ideally, each coloured 
line should trace the diagonal; however, some fluctuations are expected due to the presence of noise. 
We perform a Kolmogorov–Smirnov (KS) test~\cite{2020SciPy-NMeth} and quote the p-value (in brackets) as a measure of the consistency between each colored line and the diagonal line.
To further quantify this consistency, we can define a binomial random variable $X$ as the number of times the injected value is recovered within 
a confidence interval corresponding to a value on the horizontal axis. 
For this binomial distribution, the shape parameter $N$ equals 70 and $p$ equals the confidence corresponding to the x-axis.  
The shaded regions in Fig.~\ref{pp-plot} cover 1$\sigma$, 2$\sigma$, and 3$\sigma$ confidence intervals of the binomial probability distribution in decreasing order of opacity. 
As all the plotted curves for the different parameters fall within the 3$\sigma$ boundary and the accompanying p-values indicate good consistency with the diagonal, 
we conclude that our method is robust.

\begin{figure}
\includegraphics[width=\linewidth]{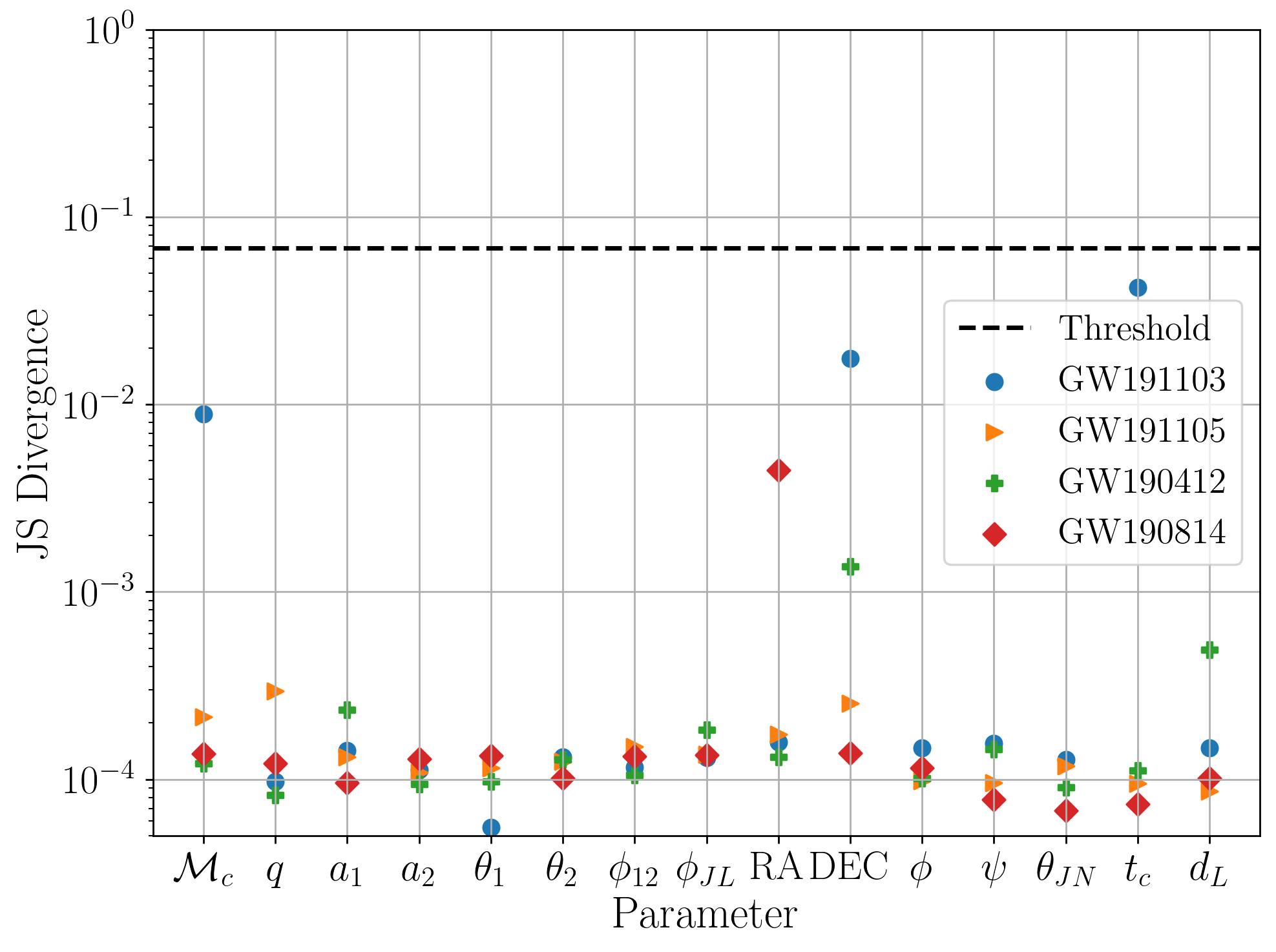}
\caption{JS divergence between the posterior distributions obtained using the regular and the relative binning methods for each source parameter of each event. 
Each event is represented by a different marker and the black dashed line shows the threshold value. The JS divergence for each parameter remains well below the 
threshold defined in the main text, indicating that the posterior distributions produced by the relative binning method and the regular method are statistically indistinguishable.}
\label{individual-js}
\end{figure}

\begin{figure}
\includegraphics[width=\linewidth]{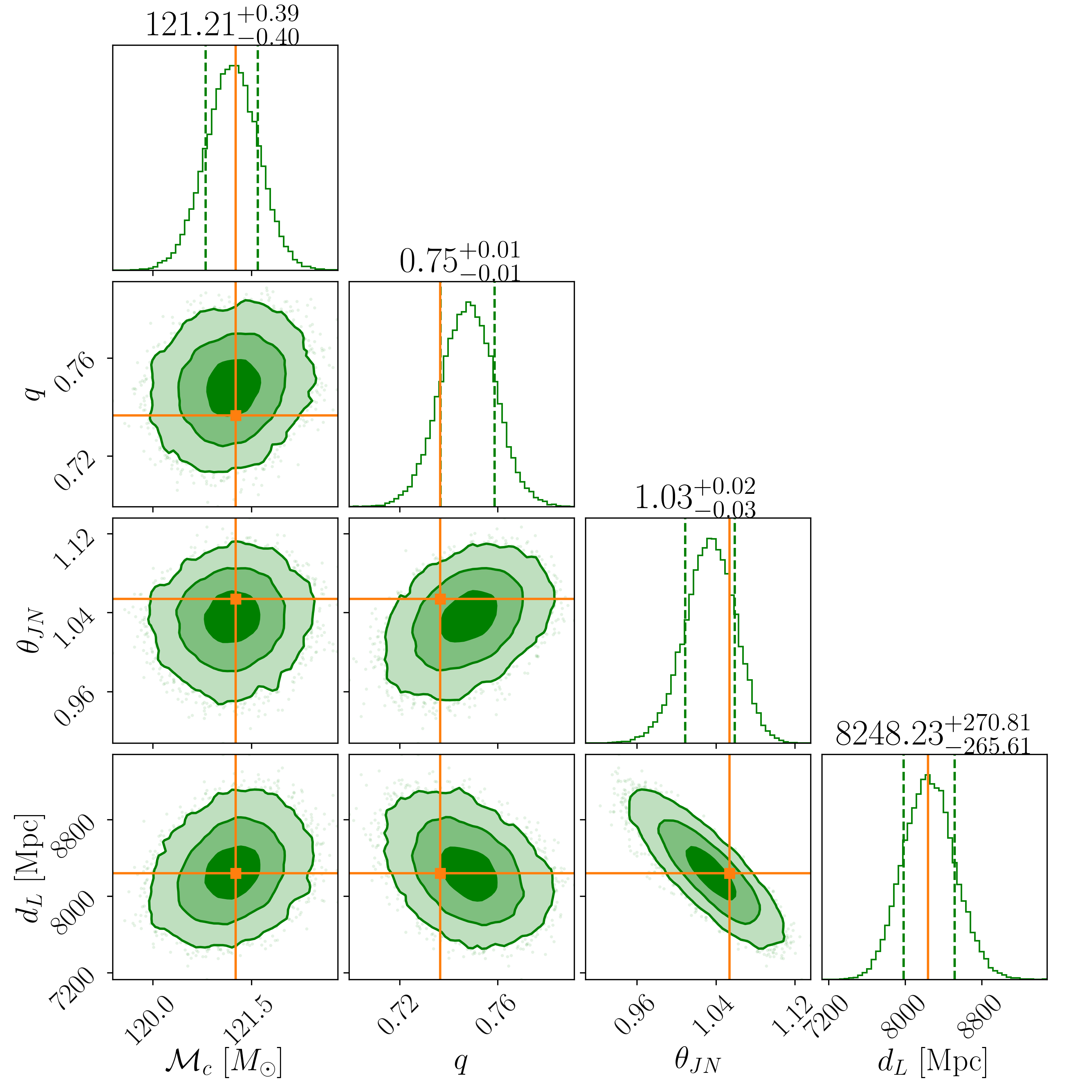}
\caption{A truncated corner plot showing the results of parameter estimation performed using relative binning on a GW200220\_061928 
signal as seen by the ET+CE network of detectors. The orange lines show the injected values and the dashed error bars represent 1$\sigma$ confidence intervals. 
The green contours correspond to $1\sigma$, $2\sigma$, and $3\sigma$ confidence, in order of decreasing opacity.}
\label{3g-event}
\end{figure}

\begin{figure}
\includegraphics[width=\linewidth]{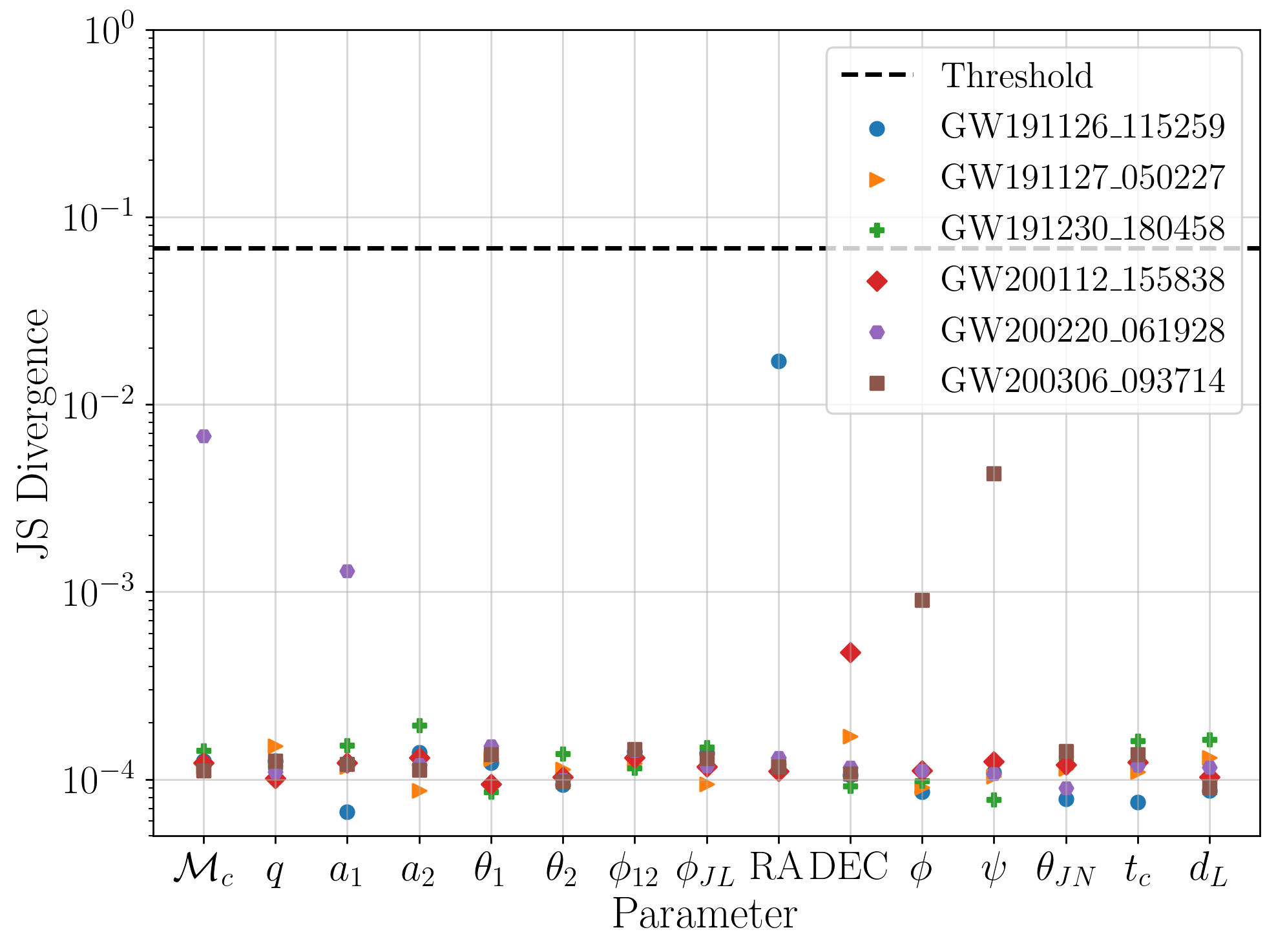}
\caption{The same as in Fig.~\ref{individual-js}, but for each of the 3G-like events. Also here the JS divergence for 
each parameter remains well below the threshold, again indicating that the posterior distributions produced by the relative binning method and the 
regular method are statistically indistinguishable.}
\label{three-g-js}
\end{figure}

Next we turn to parameter estimation on observed events. 
We choose GW190412, GW190814, GW191103, and GW191105. 
The first two events are chosen as they consist of binaries with highly asymmetric masses. This feature allows for a significant contribution from the higher 
harmonics~\cite{LIGOScientific:2020stg}, enabling us to stress-test our framework regarding HOMs.  
The remaining two events were chosen because they were part of a candidate event pair from the strong lensing searches~\cite{LIGOScientific:2023bwz, Janquart:2023mvf}. 

For GW190412 and GW19084, a truncated corner plot showing posterior distributions for selected parameters is presented in Figs.~\ref{gw190412} and \ref{gw190814}, respectively.
The green (blue) curves represent the results of parameter estimation with (without) relative binning. 
The blue curves are generated by using the posterior samples from the GWOSC data release~\cite{LIGOScientific:GWTC3DR, LIGOScientific:GWTC2.1DR}. 
Figs.~\ref{gw190412} and \ref{gw190814} show a qualitative agreement between the posteriors distributions obtained by the regular method and the relative binning method. 
Similarly, we analyzed GW191103 and GW191105 with the relative binning method, which led to comparable agreement with the regular method. 

To quantify the mismatch between the posterior samples obtained using the relative binning method to ones obtained with the regular method, we use the Jensen-Shannon 
(JS) divergence~\cite{2020SciPy-NMeth}. 
We do not perform a regular parameter estimation run due to computation cost. Instead, we reweight the posterior samples obtained using the relative binning 
method to obtain the posterior samples for the regular method as discussed in Appendix~\ref{reweight}.
We repeat this step for all four events, GW190412, GW190814, GW191103, and GW191105, and plot the JS divergence between the relative 
binning samples and the regular samples for each parameter in Fig.~\ref{individual-js}. Each event is represented by a different marker. 
To determine a heuristic threshold value for statistical indistinguishability (represented by the dashed line in Figs.~\ref{individual-js},~\ref{three-g-js}, and~\ref{median-jpe}), 
we compute the JS statistic for 2000 pairs of sets of $10^5$ samples taken from Gaussian distributions. 
We then define as heuristic threshold for indistinguishably the 99th percentile of the obtained values 
(black dashed line in Fig.~\ref{gw190412} and Fig.~\ref{gw190814}). The JS divergence for each parameter remains well below the threshold, 
indicating that the posterior distributions produced by the relative binning method and the regular method are statistically indistinguishable.

\subsubsection{3G \lowercase{Interferometers}}
\label{sss:3g-sensitivity}
Now we turn to parameter estimation for the case of 3G observatories. 
In order to test our method, we analyze a subset of GWTC-3 events as they would be seen by these detectors. 
We arbitrarily choose six events with varying total masses from the GWTC-3 catalog, specified in table~\ref{tab:gwtc-3-table}. 

\begin{widetext}
\begin{table}[h]
    \centering
    \begin{tabular}{|c|c|c|}
        \hline
        \textbf{Event} & \textbf{Total Mass ($M_{\odot}$)} & \textbf{Network SNR (ET+CE)} \\
        \hline
        GW191126\_115259 & 26.51 & 192.52 \\
        \hline
        GW191127\_050227 & 138.71 &  260.33 \\
        \hline
        GW191230\_180458 &  144.69 & 383.10 \\
        \hline
        GW200112\_155838 & 78.10 & 1103.53 \\
        \hline
        GW200220\_061928 &  282.50 & 224.91 \\
        \hline
        GW200306\_093714 & 62.10 &  326.02 \\
        \hline
    \end{tabular}
    \caption{Table of events from GWTC-3, their (maximum-likelihood, observed) total masses, and their respective network SNRs as obtained for a network of 3G detectors.}
    \label{tab:gwtc-3-table}
\end{table}
\end{widetext}

We inject the signal corresponding to the maximum likelihood parameters, as obtained from the GWOSC data release, into Gaussian noise and perform parameter estimation with 
the relative binning framework. 
Our detector network consists of a triangular ET~\cite{Punturo:2010zz, Maggiore2019ScienceTelescope, Branchesi:2023mws} with an arm length of 10 km and one L-shaped CE~\cite{Reitze:2019iox} with an arm length of 40 km.
The noise curves and detector configurations are read from~\cite{threeGdesign}. 
A representative, truncated corner plot for this analysis is shown in Fig.~\ref{3g-event} for the GW200220\_061928 event. 
The orange lines in the plot show the injected parameter values and the green dashed lines indicate $1\sigma$ confidence intervals. 
The $1\sigma$, $2\sigma$, and $3\sigma$ confidence contours are indicated by the green shading, in order of decreasing opacity. 
As can be seen from the plot, all injected parameters have been recovered well. 

To quantify the accuracy of our method for the 3G scenario, we repeat the steps which were performed in subsection~\ref{sss:hlv-sensitivity}.
We again reweight the posterior samples obtained using the relative binning method following the procedure in Appendix~\ref{reweight} and 
calculate the JS divergence between the reweighted samples and the relative binning samples. 
This process is repeated for each of the six events and the results are shown in Fig.~\ref{three-g-js}. 
Each event is represented by a different marker and the dashed black line shows the threshold value, which is calculated in the same way as in~\ref{sss:hlv-sensitivity}. 
Here too the JS divergence for each parameter remains well below the threshold, indicating that the posterior distributions produced by the relative binning method and the 
regular method are statistically indistinguishable.

\subsection{Joint \lowercase{Parameter Estimation}}
\label{subsec:jpe}

\begin{figure}
\includegraphics[width=\linewidth]{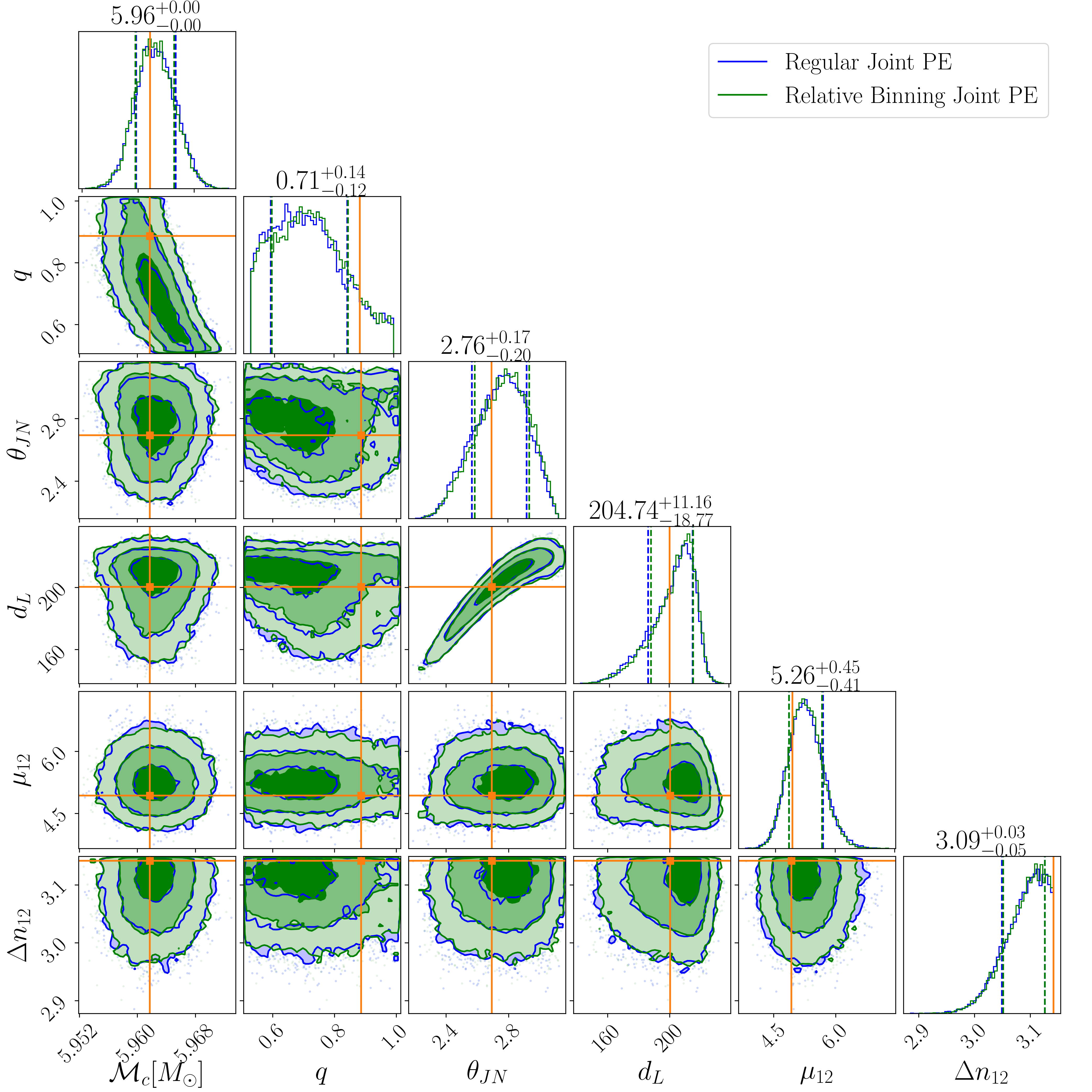}
\caption{Comparison of the posterior distribution obtained using regular parameter estimation (blue curves) and relative binning (green curves) for a simulated lensed signal. 
The orange lines show the injected values while the dashed error bars indicate a 1-$\sigma$ confidence interval.}
\label{jpe-injection}
\end{figure}

\begin{figure}
\includegraphics[width=\linewidth]{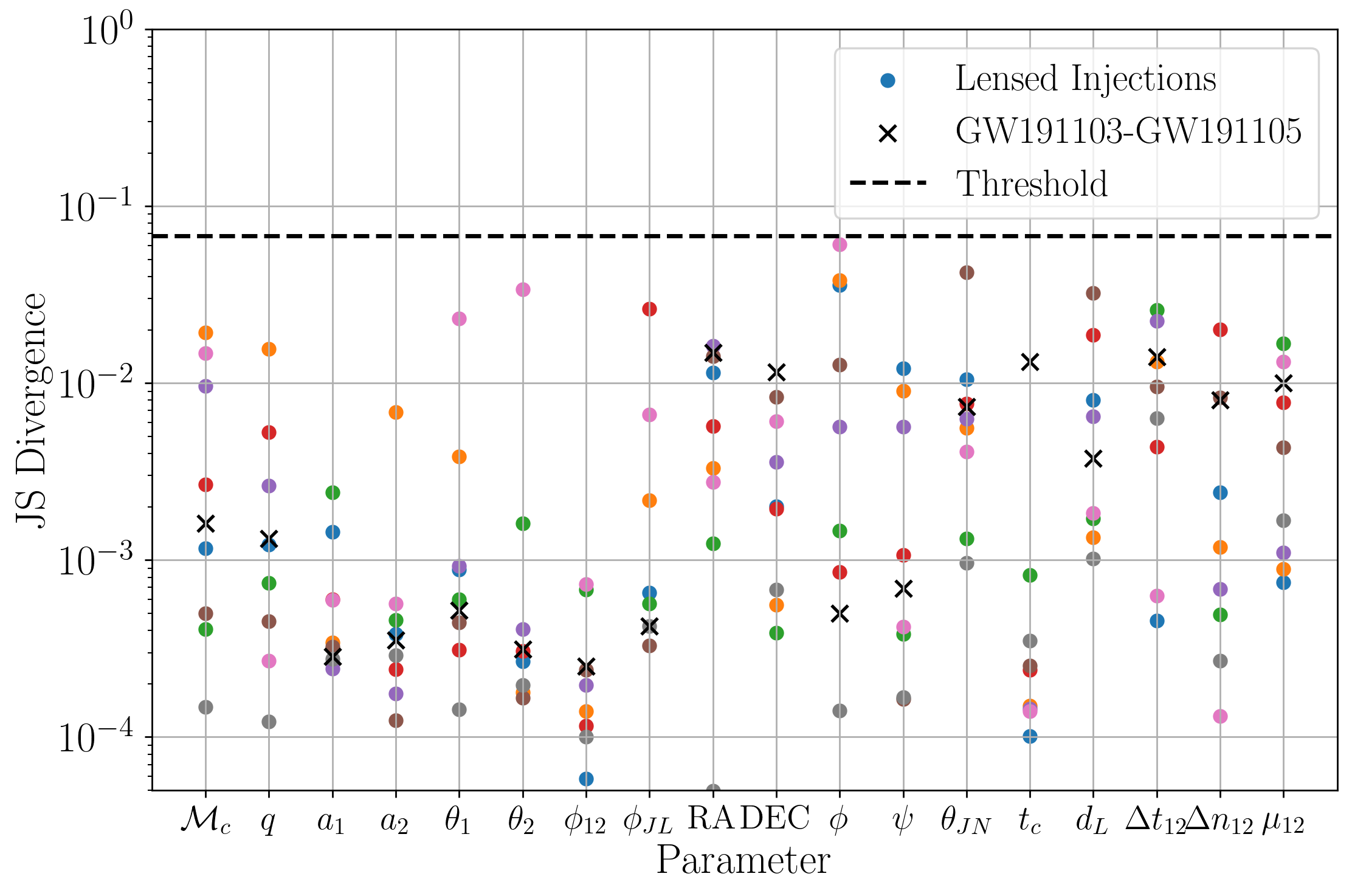}
\caption{JS divergence between the posterior distributions obtained using the regular and the relative binning methods for each source parameter for the GW191103--GW191105 event pair and for the injections. A total of 8 lensed injections have been analyzed and each one is marked by a specific coloured dot. 
The JS divergence for each parameter remains below the threshold, indicating that the posterior distributions produced by the two methods are statistically indistinguishable.}
\label{median-jpe}
\end{figure}

\begin{figure}
\includegraphics[width=\linewidth]{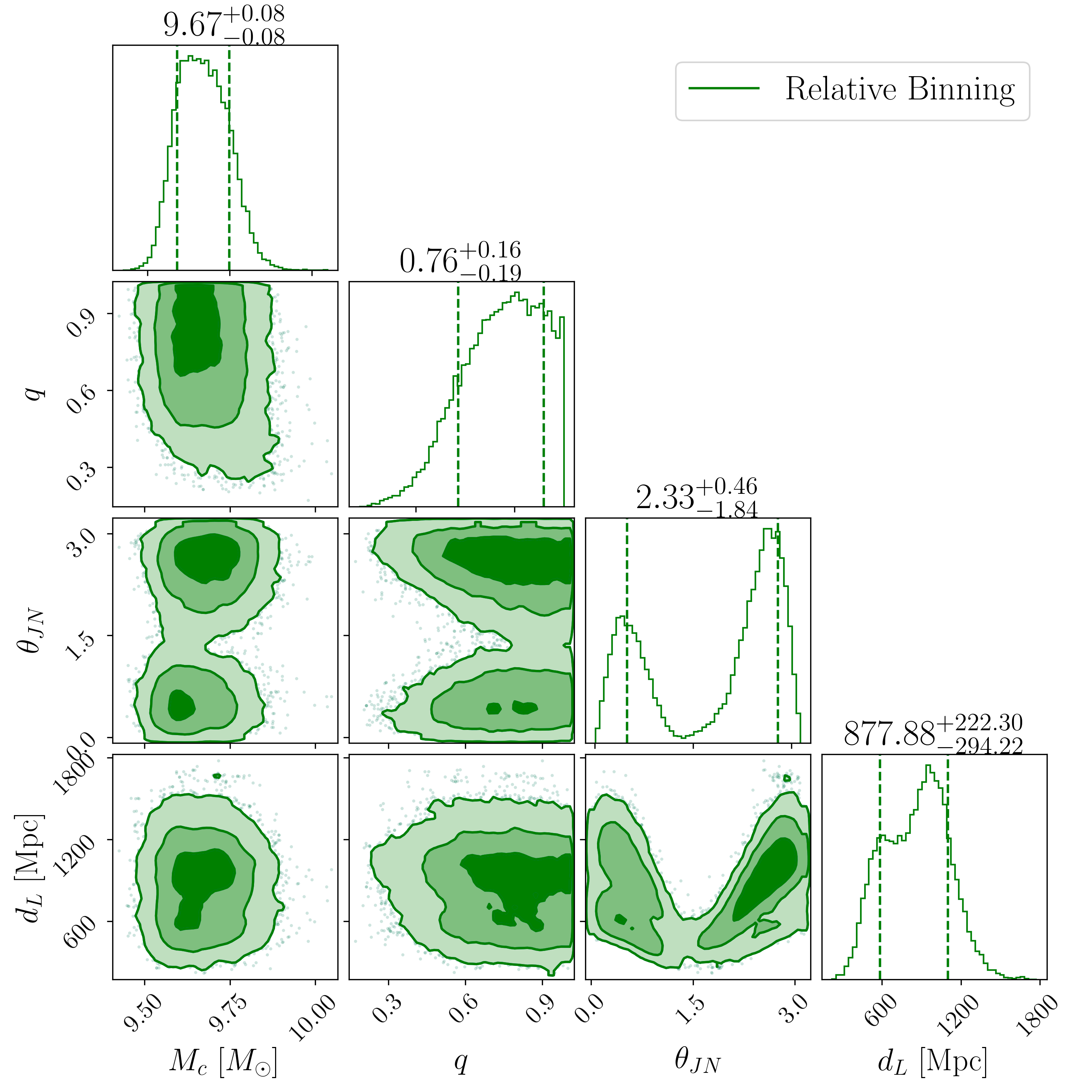}
\caption{Truncated corner plot showing the recovery of common parameters for the GW191103--GW191105 event pair.}
\label{common_params}
\end{figure}

\begin{figure}
\includegraphics[width=\linewidth]{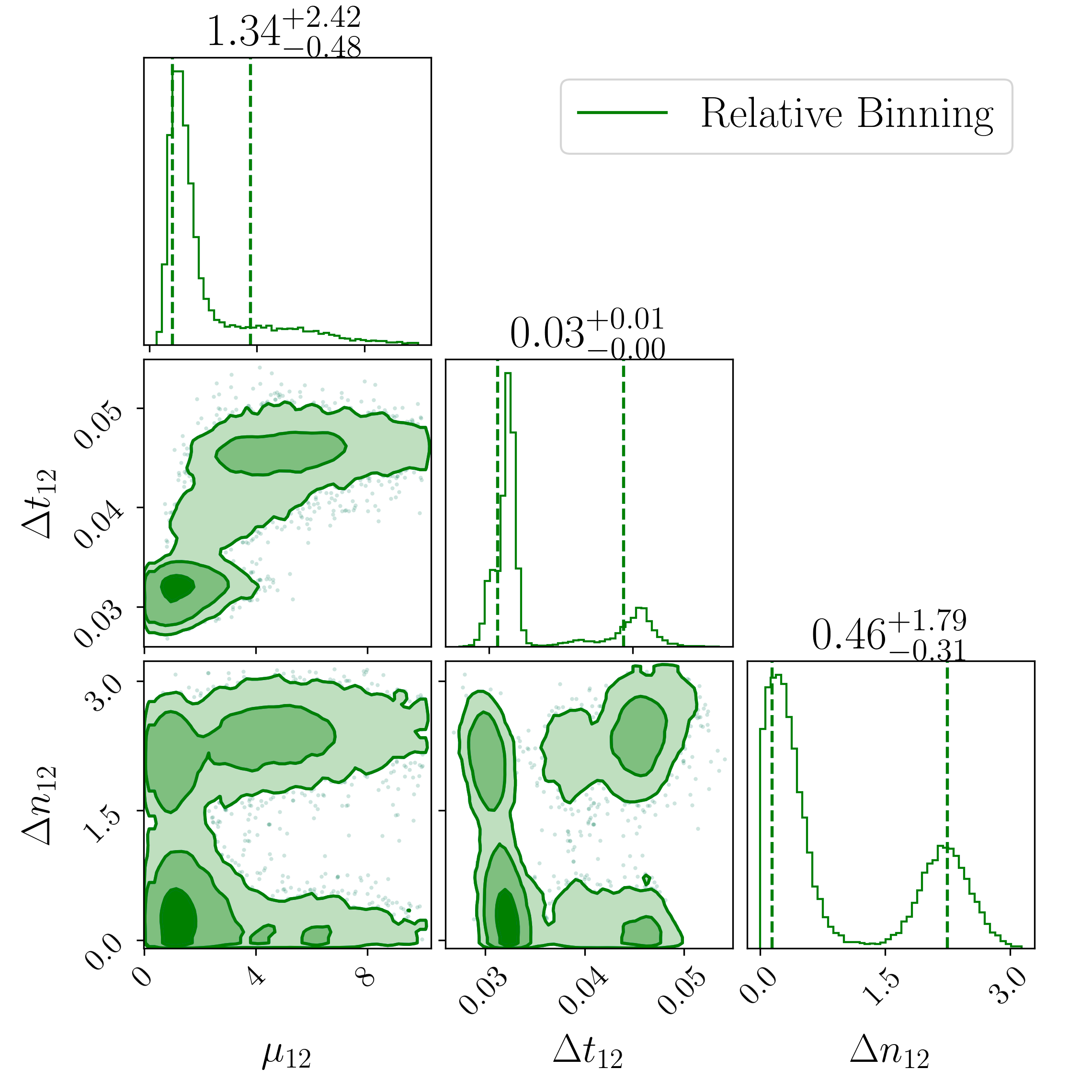}
\caption{Corner plot showing the recovery of the relative lensing parameters for the GW191103--GW191105 event pair. The time delay ($\Delta t_{21}$) posterior has been shifted by the difference between the arrival time of the GW191105 and GW191103 so that it can peak at around 0.}
\label{lensed_parameters}
\end{figure}

We analyze eight simulated lensed pairs and the GW191103--GW191105 event pair to test the relative binning framework for joint parameter estimation. 
The latter event pair is of interest as it was followed up by strong lensing searches because it showed some prototypical features of galaxy lensing, such as a mild posterior overlap and a short time delay~\cite{LIGOScientific:2023bwz, Janquart:2023mvf}. 
On the other hand, the simulated pairs were generated by sampling the relevant source parameters from the priors shown in Table~\ref{bbh_params}. 
We adjust the luminosity distance and the relative magnification of each lensed pair to ensure that the network SNR of each image is above 13.

Before delving into a detailed comparison between the relative binning and regular parameter estimation results, we present a truncated corner plot for a representative lensed pair in Fig.~\ref{jpe-injection}. 
As before, the green and blue curves show the results obtained using the relative binning and regular parameter estimation frameworks, with 
the orange lines representing the injected values of the parameters. 
The injected values are accurately recovered, and there is good agreement between methods.

For a more quantitative comparison, we compute the JS divergence between the posterior distributions obtained using the regular and relative binning methods;  
see Fig.~\ref{median-jpe}. 
Note that, for the lensed injections, we do not use reweighting as was done for the individual parameter estimation cases. 
We instead perform the regular joint parameter estimation runs with \textsc{Golum}~\cite{Janquart:2023osz} and compare the results with the ones obtained using relative binning. Each injection is represented by a different coloured dot.
All dots are below the threshold value, suggesting that the posterior distributions obtained using the regular and relative binning frameworks are statistically indistinguishable.

Now let us look at joint parameter estimation results for the GW191103--GW191105 pair. 
In this analysis, GW191103 is treated as image 1, and GW191105 as image 2. 
Fig.~\ref{common_params} shows a truncated corner plot of the common parameters between the two images, while Fig.~\ref{lensed_parameters} presents the posterior distribution of the relative lensing parameters connecting the lensing parameters of images 1 and 2.
To quantify the accuracy of our results, we follow the same reweighting steps mentioned towards the end of subsections \ref{sss:hlv-sensitivity} and \ref{sss:3g-sensitivity} in order to calculate the JS divergence between the relative binning samples and the regular samples for each parameter. 
The JS divergences are marked with black crosses in Fig.~\ref{median-jpe}, which are all below the threshold. 

Finally, we report that the log of coherence ratio~\cite{Janquart:2023osz, Lo:2021nae, Janquart:2021qov} obtained for GW191103--GW191105 is 4.506. 
This is in favor of the lensing hypothesis and comparable to the one reported by the lensing follow-up searches~\cite{Janquart:2023mvf}. 

\subsection{Speed-\lowercase{up}}
\label{subsec:speed-up}

\begin{figure}
\includegraphics[width=\linewidth]{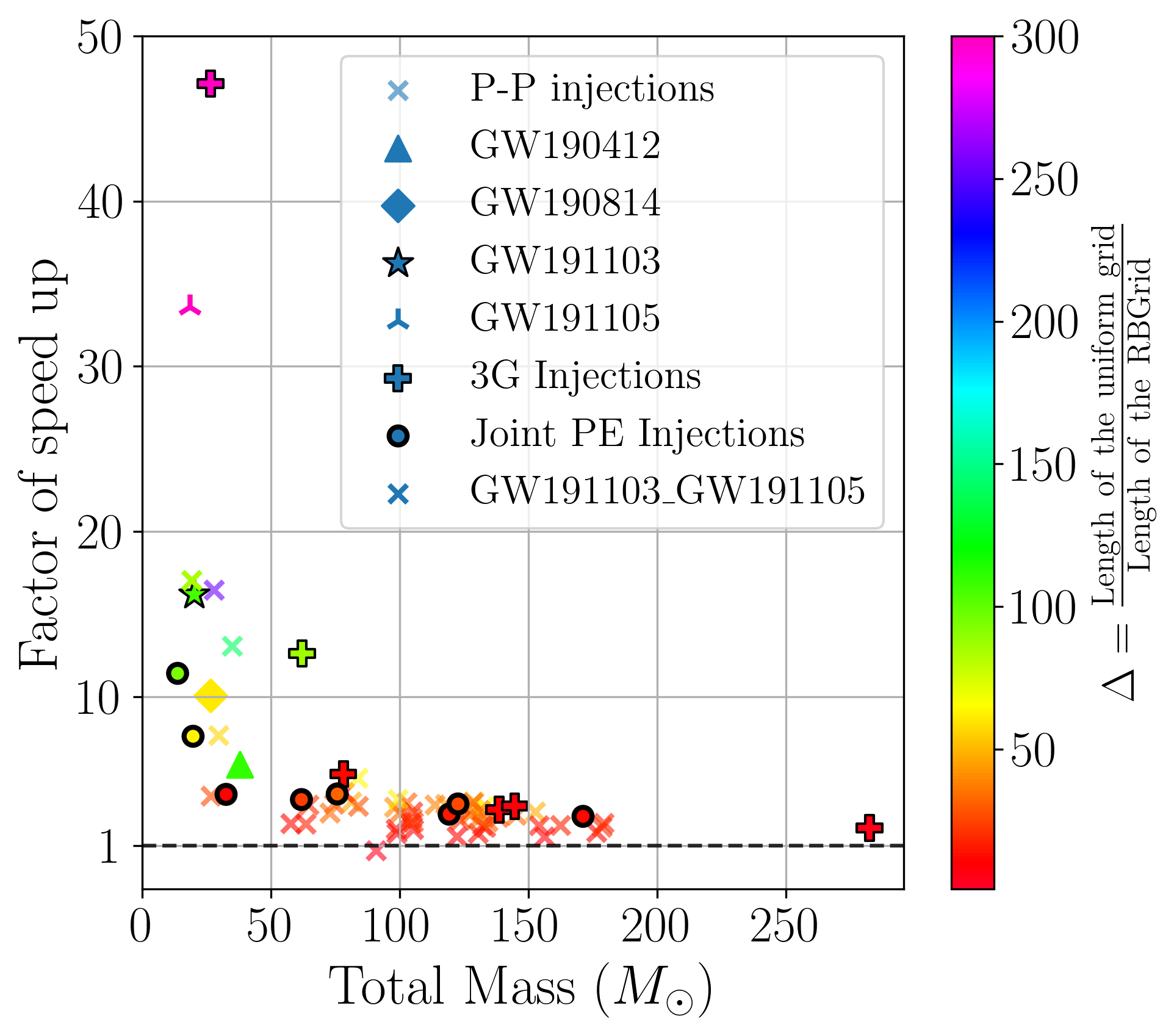}
\caption{Speed-up factors for the relative binning methods compared to regular methods. 
The circles indicate the ratio of the total sampling time taken by the regular method to the relative binning method. 
The crosses show the ratio of the average time taken to perform single likelihood evaluation by the two methods. 
The colorbar shows the ratio of the length of the uniform grid to the length of the corresponding RBGrid. 
The speed-up factor tends to be proportional to $\Delta$ and to increase with decreasing total mass.}
\label{speed-stat}
\end{figure}

The speed-up achieved by the relative binning method can be broadly attributed to two factors. 
First, every proposal waveform is generated on fewer frequency nodes -- provided by the RBGrid -- compared to the regular method, reducing the time required to generate the waveforms during sampling. 
Second, as the summary data are pre-computed, the summation in Eqs.~\eqref{eqn:dh}-~\eqref{eqn:hh} needs to be computed at fewer frequency points, reducing the number of operations.
The first point is the more significant contributor to the overall speed-up.
With this, the total speed-up for relative binning mainly depends on how coarse the RBGrid is compared to the corresponding uniform grid. The reduction in frequency nodes depends on the signal characteristics such as the duration, SNR, fiducial parameters, and also the total error chosen to generate the RBGrid.

Here we use the various parameter estimation runs done in previous sections to asses the speed-up obtained when using relative binning compared to traditional methods.
Fig.~\ref{speed-stat} shows the speed-up achieved by the relative binning method over the regular method as a function of the total mass of the systems. The colorbar shows the ratio of the length of the uniform frequency grid to the length of the corresponding RBGrid. We denote the ratio by $\Delta$.  
The circles indicate the ratio of total sampling time taken by the regular method compared to the relative binning method. 
The rest of the markers show the ratio of the average single likelihood evaluation time taken between the two methods calculated over a set of $10^4$ points. The latter is an approximation but makes it possible to 
assess the speed-up factor given by relative binning without having to do the regular parameter estimation runs. Still, for our various tests, 
the number of likelihood evaluations done for the relative binning and regular methods is in the same ballpark, so that this is a reasonable approximation to make.

We observe that the relative binning runs are up to a factor of $\sim 34$ faster for the LIGO-Virgo interferometers and up to a factor of $\sim 47$ faster for the 3G interferometers. 
The speed-up decreases as the total mass of the system increases. 
This is as expected since the grid can be made the coarsest for the inspiral part of the signal, which is less prominent for higher mass systems. 
However, within a given total mass regime, some fluctuations in the speed-up can happen due to other signal characteristics.   
The 3G detectors have a lower starting frequency, meaning they can observe a given signal for a longer period compared to the LIGO-Virgo interferometers. 
Therefore, one might expect relative binning to show a larger speed-up in the 3G scenario compared to similar total systems as seen in LIGO-Virgo, and 
this effect is partially visible in Fig.~\ref{speed-stat}. However, the 3G injections considered for the analysis have a considerably higher SNR than the corresponding 
P-P injections for the LIGO-Virgo interferometers (blue cross markers). A higher SNR will increase the size of the coarse grid and decrease the speed of the relative binning method.

For one of the P-P injections, with a total mass of $\simeq 95 M_{\odot}$, the speed-up achieved 
by the relative binning method is below 1 (see Fig.~\ref{speed-stat}). 
This can be explained by the large network SNR ($\simeq 100$) of this particular event. 
In addition, the median of the SNR for the set of P-P injections is $59.44$  which is $\simeq 2.2$ 
times larger compared to the loudest GW event observed to date \cite{LIGOScientific:GWTC3DR}. 
Since a large value of SNR increases the size of the coarse grid, the factor of speed-up reported by the P-P injections is on the conservative side.  

Joint parameter estimation reports a similar speed-up as individual parameter estimation. Here we note that the regular joint parameter estimation framework used 
in this work also rescales the first waveform to obtain the second one, by-passing the complete waveform generation step. 
However, this is not the case in all lensing analysis methods, and when joint parameter estimation methods generate the two waveforms independently, 
we expect a larger speed-up factor for relative binning.

\section{C\lowercase{onclusions and future work}}
\label{sec:conclusions}

In this work, we have presented a relative binning framework capable of doing parameter estimation analyses with waveforms including precession and higher-order 
modes. Our relative binning codebase can be found in~\cite{janquart:2022}.
This marks the first instance where the relative binning method is used to perform parameter estimation with waveforms containing HOMs and precession.
Additionally, we have extended the method to perform joint parameter estimation for strongly-lensed signals, representing the first time the relative binning method is employed in such a context. 
By analyzing a large set of injections and a select set of observed GW signals, we have demonstrated that our method can perform parameter estimation with an accuracy comparable 
to the regular method, but with a significant speed-up. 
The latter decreases as the total mass of the system increases since the inspiral of the signal is the region where we can make the coarsest grid. 
However, low-mass signals are the most computationally heavy to analyze and the larger speed-up for those signals is an important feature.

To get an even more rapid parameter estimation tool based on relative binning, some upgrades can be brought to our framework. For example, currently, the $h^L_{l, m}$ terms are calculated using the \textsc{C/C++} based routine \texttt{LALSimulation}~\cite{lalsuite-ref}, and the $C_{l, m}$ coefficients are calculated using a Python code accelerated by $\texttt{Numba}$~\cite{lam2015numba}.
While the latter can approach the speed of \textsc{C/C++}~\cite{lam2015numba}, implementing the calculation of $C_{l, m}$ in \texttt{LALSimulation} could result in additional speed-up. This would require modifying \texttt{LALSimulation} itself as it is currently not capable of computing the $C_{l, m}$ terms on a selected frequency grid.

The performance of our method depends on the choice of the fiducial waveform. 
In this work, we have assumed that the fiducial waveform is either (a) the same as the injected waveform for simulations, or (b) the maximum likelihood waveform for real events. 
However, the method can be seeded in a more realistic way by using the best-fitting template reported by the templated-based searches or maximum likelihood estimator routines can be used to determine the fiducial waveforms' parameters~\cite{Srivastava:2018wvy, Usman:2015kfa, Sachdev:2019vvd, Aubin:2020goo, Finstad:2020sok}.  
We plan to work on the choice of the fiducial waveform and understanding how it can impact the performance of our method in the future. This is an important step if one wants to use relative binning techniques on real data, where we do not know parameters in advance. 

The detection rates of GWs and consequently the amount of computational resources needed to perform parameter estimation will increase with detector upgrades. 
In order to achieve various scientific goals, there is a need for a much faster, and at the same time precise, framework to estimate parameters than what is offered 
by traditional methods. This work represents an important step in this direction, as it provides a methodology capable of performing various analyses with 
waveform models that are physically reasonably complete, in a much accelerated way.

\begin{acknowledgments} 
The authors would like to thank Anna Puecher, Chinmay Kalaghatgi, Marc Van Der Sluys, Melissa Lopez, Stefano Schmidt, and Tomasz Baka for the useful discussion. 
We also thank Mick Wright for carefully re-reading the manuscript.
H.N., J.J., Q.M., K.H., and C.V.D.B.~are supported by the research programme 
of the Netherlands Organisation for Scientific Research (NWO). 
This material is based upon work supported by NSF's LIGO Laboratory which is a major facility fully funded by the National Science Foundation.
This research has made use of data, software and/or web tools 
obtained from the Gravitational Wave Open Science Center (https://www.gw-openscience.org), a 
service of LIGO Laboratory, the LIGO Scientific Collaboration and the Virgo Collaboration. 
LIGO is funded by the U.S. National Science Foundation. Virgo is funded by the French
Centre National de Recherche Scientifique (CNRS), the 
Italian Istituto Nazionale della Fisica Nucleare (INFN) 
and the Dutch Nikhef, with contributions by Polish and Hungarian institutes.
\end{acknowledgments}

\appendix
\section{Reweighting of the relative binning posterior samples}
\label{reweight}
We reweight the posterior samples obtained using relative binning to match the samples obtained with the usual likelihood to test the method's efficiency. 
Let us denote the likelihood evaluated for the posterior samples produced by the relative binning 
method by $\mathcal{L}^{\text{RB}}(\boldsymbol{\Theta})$, where $\boldsymbol{\Theta}$ stands for the 
posterior samples; and the probability for $\boldsymbol{\Theta}$ according to the relative binning 
samples by $p^{\text{RB}}(\boldsymbol{\Theta})$. To perform the reweighting, for each posterior sample $\boldsymbol{\Theta}$, we compute the likelihood according to the usual method, which we denote by $\mathcal{L}^{\text{RG}}(\boldsymbol{\Theta})$. 
Using $\mathcal{L}^{\text{RB}}(\boldsymbol{\Theta})$, $\mathcal{L}^{\text{RG}}(\boldsymbol{\Theta})$, 
we can reweight $p^{\text{RB}}(\boldsymbol{\Theta})$ to obtain the probability 
for $\boldsymbol{\Theta}$ according to the regular method:
\begin{equation}
p^{\text{RG}}(\boldsymbol{\Theta})  = \frac{\mathcal{L}^{\text{RG}}(\boldsymbol{\Theta})}{\mathcal{L}^{\text{RB}}(\boldsymbol{\Theta})}p^{\text{RB}}(\boldsymbol{\Theta})\,.
\end{equation}
\bibliography{references}

\end{document}